\documentclass[twocolumn,secnumarabic,amssymb,nobibnotes,aps,prd,superscriptaddress,nofootinbib]{revtex4-2}

\usepackage[utf8]{inputenc} % allow utf-8 input
\usepackage[T1]{fontenc}    % use 8-bit T1 fonts

\usepackage{url}            % simple URL typesetting
\usepackage{booktabs}       % professional-quality tables
\usepackage{amsfonts}       % blackboard math symbols
\usepackage{nicefrac}       % compact symbols for 1/2, etc.
\usepackage{microtype}      % microtypography
\usepackage{amsmath}        % more mathematics!
\usepackage[capitalise]{cleveref}       % more clever refs
\usepackage{chemformula}    % chemical formulas
\usepackage[separate-uncertainty=true,exponent-product=\cdot]{siunitx}        % nicely typeset units
\DeclareSIUnit\angstrom{\text{\AA}} % angstrom is not a SI unit :clown_face:
\usepackage{graphicx}
\usepackage{xcolor}
\usepackage{mathtools}
\usepackage{todonotes}
\usepackage{xspace}
\usepackage{tikz}
\usetikzlibrary[arrows.meta,bending]
\usetikzlibrary{positioning}

\usepackage[normalem]{ulem}

\newcommand{\scare}[1]{\lq#1\rq} 
\newcommand{\etal}{et~al.}

\newcommand{\symset}[1]{\mathcal{#1}}  % symbol of a set
\newcommand{\tk}{\boldsymbol{\kappa}}
\newcommand{\J}{{\bf J}}
\newcommand{\Jfull}{\J_\text{Hardy}}

\newcommand{\Jconv}{\J_{\text{convective}}}
\newcommand{\Jmpnn}{\J_{\text{Hardy}}^{\text{MPNN}}}

\newcommand{\Jfan}{\J_{\text{Fan}}}

\newcommand{\dur}[3]{\frac{\partial U_{#1}}{\partial \R_{#2#3}}}
\newcommand{\indur}[3]{\partial U_{#1} / \partial \R_{#2#3}}

\newcommand{\nbh}[1]{\symset{N}(#1)}

\newcommand{\interactions}{M}
\newcommand{\cutoff}{r_{\text{c}}}
\newcommand{\effcutoff}{\cutoff^{\text{eff}}}

\newcommand{\graph}{\mathcal{G}}
\newcommand{\vertices}{\mathcal{V}}
\newcommand{\edges}{\mathcal{E}}

\newcommand{\R}{{\bf R}}

\newcommand{\Rm}{{\bf R}^{\text{MIC}}}

\newcommand{\V}{\dot{\bf R}}

\newcommand{\Bary}{{\bf B}}
\newcommand{\intd}{\text{d}}

\newcommand{\defdas}{\eqqcolon}

\newcommand{\curlyset}[2]{\{\, #1\, :\, #2 \, \}}

\newcommand{\magnitude}[1]{|#1|}

\newcommand{\Rsc}{\symset{R_{\text{sc}}}}

\newcommand{\Rall}{\symset{R_{\text{all}}}}

\newcommand{\Runf}{\symset{R_{\text{unf}}}}
\newcommand{\Jpot}{\J_\text{pot}}
\newcommand{\Junf}{\J_\text{pot}^{\text{unfolded}}}
\renewcommand{\Jconv}{\J_\text{conv}}
\renewcommand{\Jfull}{\J}
\renewcommand{\Jfan}{\J_\text{pot}^{\text{local}}}
\renewcommand{\Jmpnn}{\J_{\text{pot}}^{\text{semi-local}}}

\newcommand{\abr}[1]{\uppercase{#1}}

\newcommand{\nns}{\abr{nn}s\xspace}
\newcommand{\mpnn}{\abr{mpnn}\xspace}

\newcommand{\mlp}{\abr{MLP}\xspace}
\newcommand{\mlps}{\abr{MLP}s\xspace}

\newcommand{\ffs}{\abr{ff}s\xspace}
\newcommand{\ff}{\abr{ff}\xspace}
\newcommand{\ad}{\abr{ad}\xspace}
\newcommand{\gk}{\abr{gk}\xspace}

\newcommand{\md}{\abr{md}\xspace}
\newcommand{\dft}{\abr{dft}\xspace}
\newcommand{\pes}{\abr{pes}\xspace}

\newcommand{\mic}{\abr{mic}\xspace}

\definecolor{yellow1}{HTML}{FADF63}
\definecolor{yellow2}{HTML}{E6AF2E}
\definecolor{light_red}{HTML}{FE404A}

\definecolor{extra_0}{HTML}{66c2a5}
\definecolor{extra_1}{HTML}{ffd92f}
\definecolor{extra_2}{HTML}{8da0cb}
\definecolor{extra_3}{HTML}{e78ac3}
\definecolor{extra_4}{HTML}{a6d854}
\definecolor{extra_5}{HTML}{fc8d62}
\definecolor{accent_2}{HTML}{006C66}

\definecolor{signalred}{HTML}{EE4B2B}

\newcommand{\mrc}[1]%
{\todo[inline,backgroundcolor=orange,size=\small, bordercolor=white]{{\bf MR:} #1}}

\newcommand{\SM}{Supp. Mat.\xspace}

\usepackage{natbib}
\newcommand{\nomadaff}{\affiliation{The NOMAD Laboratory at the FHI of the Max-Planck-Gesellschaft and IRIS Adlershof of the Humboldt Universit{\"a}t zu Berlin, Germany}}

\begin{document}

\title{Heat flux for semi-local machine-learning potentials}

\begin{abstract}
    The Green-Kubo (GK) method is a rigorous framework for heat transport simulations in materials.
    However, it requires an accurate description of the potential-energy surface and carefully converged statistics.
    Machine-learning potentials can achieve the accuracy of first-principles simulations while allowing to reach well beyond their simulation time and length scales at a fraction of the cost.
    In this paper, we explain how to apply the GK approach to the recent class of message-passing machine-learning potentials, which iteratively consider semi-local interactions beyond the initial interaction cutoff. 
    We derive an adapted heat flux formulation that can be implemented using automatic differentiation without compromising computational efficiency.
    The approach is demonstrated and validated by calculating the thermal conductivity of zirconium dioxide across temperatures.
\end{abstract}

\author{Marcel F. Langer}
\email[Corresponding author: ]{mail@marcel.science}
\affiliation{Machine Learning Group, Technische Universit{\"a}t Berlin, 10587 Berlin, Germany}
\affiliation{BIFOLD -- Berlin Institute for the Foundations of Learning and Data, Berlin, Germany}
\nomadaff

\author{Florian Knoop}
\email[Corresponding author: ]{florian.knoop@liu.se}
\affiliation{Theoretical Physics Division, Department of Physics, Chemistry and Biology (IFM), Linköping University, SE-581 83 Linköping, Sweden}
\nomadaff

\author{Christian Carbogno}
\nomadaff

\author{Matthias Scheffler}
\nomadaff

\author{Matthias Rupp}
\nomadaff
\affiliation{Department of Computer and Information Science, University of Konstanz, 78464 Konstanz, Germany}
\affiliation{Materials Research and Technology Department, Luxembourg Institute of Science and Technology (LIST), Belvaux, Luxembourg}

\date{\today}%

    \maketitle

The thermal conductivity tensor $\tk$ describes the ability of a material to conduct heat when exposed to a temperature gradient.
Its computational prediction is of great interest for the design of novel high-performance materials which are needed, for example, as thermal barrier coatings in engines~\cite{ecl2008t}, or thermoelectrics for waste heat recovery~\cite{st2008t}.
Such materials often feature complex structure and strongly anharmonic potential-energy surfaces (\pes)~\cite{zldk2014t,kpsc2020t}. This implies the need to evaluate $\tk$ with the Green-Kubo (\gk{}) method~\cite{o1931At,o1931Bt,g1952t,k1957t,kyn1957t}.

In the \gk{} approach, $\tk$ is expressed in terms of the integral of the autocorrelation function of the instantaneous heat flux $\J(t)$ as observed in equilibrium molecular dynamics (\md{}) simulations,
\begin{align}
    \tk (T, p)
    = \frac{1}{k_{\rm B} T^2 V }
    \lim_{t \rightarrow \infty}
    \int_0^t \,
    \intd \tau\,
    \left\langle
    \J (\tau) \otimes \J (0) 
    \right\rangle_{T,p}
    ~,
    \label{eq:Kubo1}
\end{align}
where $k_{\rm B}$ is the Boltzmann constant, $V$ the simulation cell volume, and $\left\langle\cdot\right\rangle_{T,p}$ denotes an ensemble average at temperature $T$ and pressure $p$.

High-accuracy \md simulations can be performed using density-functional theory (\dft) when the exchange-correlation approximation is reliable~\cite{thxy2022p}. For the evaluation of \cref{eq:Kubo1}, this approach~\cite{mub2015t,crs2017t} suffers from its numerical costs which limits the system sizes and time scales that can be treated, and therefore requires additional denoising and extrapolation approaches~\cite{crs2017t,meb2020t,kcs2022t}.
The alternative, so far, was the use of semi-empirical force fields (\ffs)~\cite{g2011p}.
Here, the interatomic interactions are described by a physically-motivated analytical equation that includes free parameters which are fitted to experimental or \emph{ab initio} results.
This classical \ff approach has been very successful, as it enables a proper consideration of the ensemble averages needed in \cref{eq:Kubo1}. However, the restricted flexibility of \ffs may limit their generality and ability to model novel materials.

A new, more general class of \ffs is the family of machine-learning potentials (\mlps) which leverage techniques like neural networks (\nns)~\cite{wd2004q,lgs2004q,bp2007q,bpkc2010q,uctm2021q,pt2021q}.
\mlps offer, in principle, unrestricted flexibility, but are limited to the mechanisms and information that are provided by their training data.
In local \mlps, linear scaling with system size is achieved by using the short-ranged nature of chemical bonding~\cite{pk2005p} to 
decompose the total energy into contributions that only depend on local atomic environments.
However, a strict locality assumption limits the flexibility and therefore accuracy of such \mlps.
Some \ffs therefore include explicit long-range electrostatic and van der Waals interactions~\cite{rg1991p,hd2001p,cddw2009p,shgd2016p,kfgb2021Bq}.
Semi-local \mlps{}~\cite{gsvd2017q,sktm2017q,sstm2018q,um2019q,kgg2020q,kbg2021q,ucsm2021q,bmsk2022q,bkoc2022a,bbkc2022a,blcd2022a,fum2022a} build up longer-range correlations iteratively from local ones through message-passing mechanisms, thereby preserving linear scaling with system size. They have recently emerged as an alternative to strictly local \mlps and have shown promising performance in benchmark settings and first applications~\cite{bmsk2022q,bkoc2022a,sggm2022q,ustm2022a,co2022a}.

While local \mlps{} have been used to investigate thermal transport via \gk{}~\cite{sdbb2012q,mcld2020q,knys2019q,lqzg2021q,lll2020q,llll2020q,qpwy2019q,vkjk2021q}, more recent semi-local \mlps{} have not yet been applied, partially because a heat flux formulation that incorporates message-passing mechanisms was lacking.
In this work, we fill that gap and extend the \gk{} approach to semi-local potentials.
To this end, we derive a formulation of the heat flux that explicitly accounts for semi-local interactions, finding that the resulting thermal conductivity significantly differs from a purely local form.
While the computation of this heat flux scales quadratically with system size, we show that an alternative yet equivalent form based on an extended auxiliary system can be introduced, leading to overall linear scaling and straightforward practical implementation of the approach via automatic differentiation (\ad{})~\cite{griewank2008,bprs2017m}.
Using the SchNet message-passing neural network (\mpnn{}) \cite{sktm2017q,sstm2018q}, we demonstrate the accuracy and feasibility of large-scale semi-local \mlp{} thermal conductivity calculations for zirconia (ZrO$_2$), an oxide known 
for its strongly anharmonic \pes~\cite{fpf2001t,clws2014t}. 

For any potential function $U(\{\R_J\})$ that can be decomposed into atomic contributions $U = \sum_I U_I(\{\R_J\})$, where $\{\R_J\}$ denotes the set of all atomic positions $\R_J$, the heat flux is given by the classical equivalent of a formula by Hardy~\cite{h1963t}, which we re-derive in the \SM to explicitly account for periodic boundary conditions.

This yields the full \scare{Hardy} heat flux
\begin{align}
    \Jfull &= \sum_{\substack{I \in \Rsc \\ J \in \Rall}} \left(\R_{JI} \left(\dur{I}{J}{} \cdot \V_J\right) \right) 
        + \sum_{I \in \Rsc} E_I \V_I \label{eq:J_general}\\
        &\defdas \Jpot + \Jconv\, , \label{eq:J_pot+conv}
\end{align}
where $\V_I$ is the velocity of an atom, and $E_I = U_I + \frac{1}{2} m_I \V_I$ is the total energy per atom. For atom-pair vectors, we adopt the convention $\R_{IJ} = \R_J - \R_I$. $\Rsc$ indicate the atoms in the simulation cell, while $\Rall$ enumerates the full, infinite, bulk system.
The nomenclature for heat flux contributions and the relation of \cref{eq:J_general} to \dft formulations are further discussed in the \SM

As this work considers \ffs and \mlps that explicitly define atomic potential energies $U_I$, total atomic energies $E_I$ and consequently $\Jconv$ can be computed in a straightforward manner.
We therefore only discuss the more involved computation of $\Jpot$ in the following, whereas the heat flux used for calculating $\tk$ is always equivalent to the full flux given by \cref{eq:J_general}.

Evaluating $\Jpot$ requires disentangling the contributions of every atom, including those in the bulk, to every atomic potential energy $U_I$.
This can be challenging for non-pairwise, many-body potentials, leading to the development of specialized expressions for different \ffs~\cite{c2006t,at2011t,tno2008t,fpdh2015t,smko2019t,bbw2019t}, many of which were recently unified and shown to be equivalent to \cref{eq:J_general} by Fan~\etal~\cite{fpdh2015t}.
Their work is based on the insight that translational invariance requires that the potential is computed only from atom-pair vectors $\R_{IJ}$, which provides a convenient basis to separate the computation of each $U_I$ into distinct sets of inputs.

Combined with the introduction of an interaction cutoff radius $\cutoff$ and atomic neighborhoods $\nbh{I} = \curlyset{\R_J}{\magnitude{\R_{IJ}} \leq \cutoff, \R_J \in \Rall}$, this leads to the notion of a \emph{local} potential $U_I = U_I(\curlyset{\R_{IJ}}{J \in \nbh{I}})$ and a corresponding local formulation of $\Jpot$,
\begin{equation}
    \Jfan = \sum_{I \in \Rsc} \sum_{J \in \nbh{I}} \R_{JI} \left( \dur{I}{I}{J} \cdot \V_J \right) \, . \label{eq:J_fan}
\end{equation}
As each $\R_{IJ}$ only contributes to one $U_I$, derivatives of $U$ naturally separate into atomic contributions $\indur{I}{I}{J} = \indur{}{I}{J}$. The resulting expression can be implemented efficiently with \ad, as detailed in the \SM

While being exact for \emph{local} potentials, this formulation of the heat flux does not apply to the \emph{semi-local} case. In such potentials, longer-range interactions are introduced \emph{without} explicitly increasing the cutoff $\cutoff$ by building them up iteratively:
Neighboring atoms are allowed to exchange information for a fixed number of iterations $\interactions$~\cite{gsvd2017q}.
Neighboring environments up to an effective cutoff radius $\effcutoff = \interactions \cutoff$ therefore become correlated;
atomic potential energies $U_I$ acquire a dependence on atom-pair vectors \emph{outside} of their immediate neighborhoods $\nbh{I}$, rendering \cref{eq:J_fan} inapplicable.

To see this, we employ a description in terms of a graph $\graph$, where vertices $\vertices$ are identified with atoms $I$ in the simulation cell, and connected via edges $\edges$ labelled by atom-pair vectors $\R_{IJ}$ if they lie within $\nbh{I}$.
Semi-local \mlps{} then act on this graph by propagating information between vertices (see \SM{}). Interactions outside of the simulation cell are therefore mapped back into it, and explicit replicas are not constructed.

Assuming that $\effcutoff$ is chosen such that the minimum image convention (\mic) is applicable, $\Jpot$ in \cref{eq:J_general} can be rewritten (see \SM) as
\begin{align}
    \Jmpnn =\!\!\!\!\!\! \sum_{\substack{I \in \vertices \\ J \in \vertices \\ K \in \nbh{J}}}\!\!\!\! \Rm_{JI}\! \left( \Bigl(\dur{I}{K}{J} - \dur{I}{J}{K} \Bigr) \cdot \V_J \right) \, , \label{eq:J_graph}
\end{align}
generalizing $\Jfan$ to semi-local \mlps{}. In the case of $\interactions = 1$, this form reduces to \cref{eq:J_fan}.

\begin{figure}
  \centering
  \includegraphics[scale=0.6]{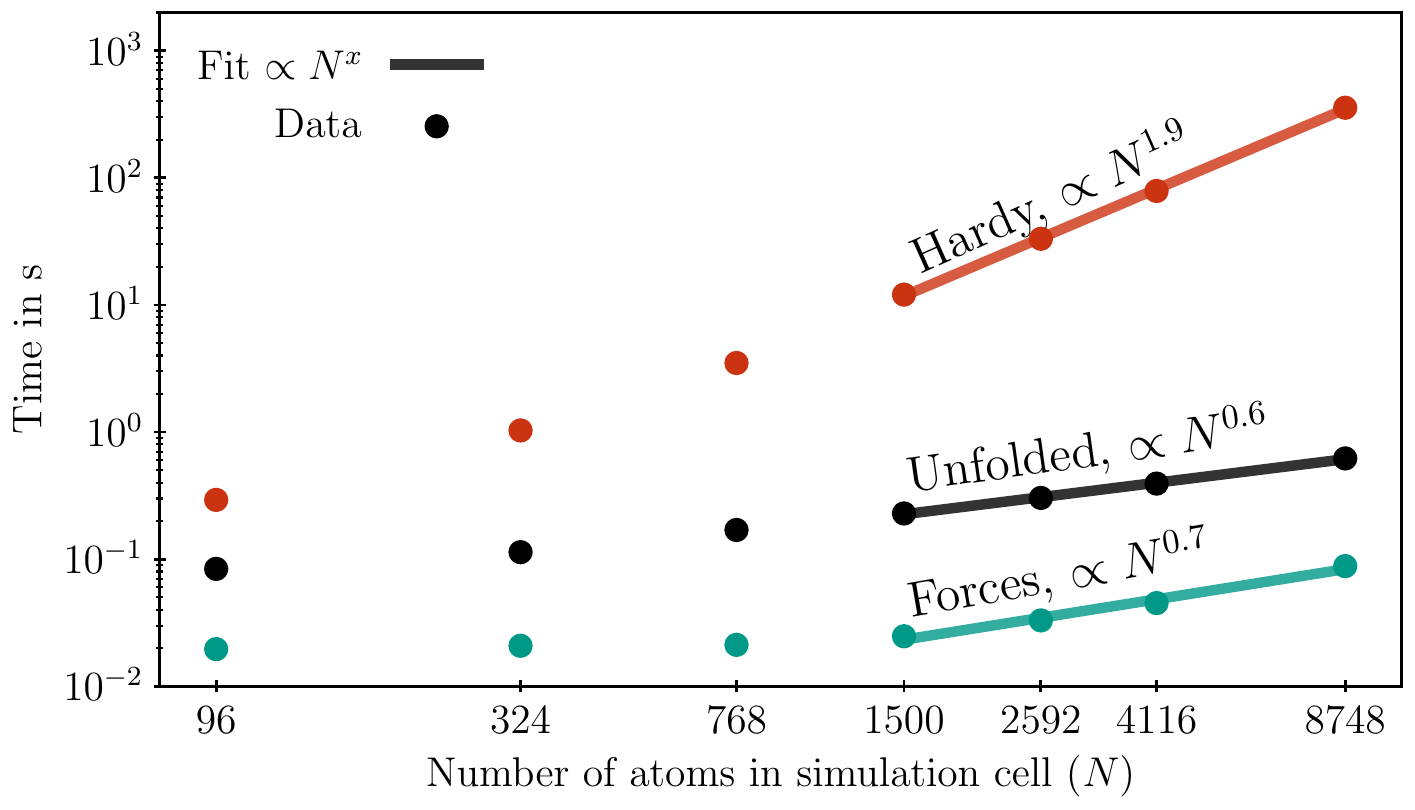}
  \caption{Computation time per timestep for different system sizes $N$ for zirconia, evaluating a SchNet \mpnn{}, for different heat flux formulations on a single Tesla Volta V100 32GB GPU. To estimate the asymptotic scaling, a function proportional to $N^x$ has been fitted to the results for large $N$. Note that on this setup with limited memory, the truly asymptotic limit cannot be reached.
}
  \label{fig:heat_flux_timings}
\end{figure}

\Cref{eq:J_graph} reflects the standard construction of semi-local \mlps{}; a double sum over all atoms is required and its evaluation formally scales quadratically with system size. As shown in~\cref{fig:heat_flux_timings}, a direct implementation of this form is therefore impractical. While force predictions for a semi-local \mlp{} based on the SchNet architecture~\cite{sktm2017q,sstm2018q} remain below \SI{100}{ms} for all system sizes investigated, the unoptimized calculation of the heat flux dominates the computational cost by several orders of magnitude at the system sizes required for the \gk{} method.

If the analytical form of $U_I$ were known, a lower-scaling evaluation of the heat flux might be accessible by deriving and implementing analytical derivatives.
Modern \mlps{}, however, typically rely on \ad{}~\cite{griewank2008,bprs2017m} for efficiently computing derivatives without requiring detailed information on the functional form of the \mlp{}.

To take advantage of this, we now derive an adapted form of the heat flux that preserves the implicit treatment of interactions beyond local environments to retain the computational efficiency of semi-local \mlps{}, while explicitly attributing all contributions to $U_I$ to bulk positions for $\Jpot$ in \cref{eq:J_general}.
This is achieved by constructing an extended simulation cell that explicitly includes all replicas that interact with atoms in the simulation cell, inspired by previous approaches which did not consider \ad{}~\cite{tpm2009t,khc2012t}. The graph representation is then constructed \emph{without} periodic boundary conditions, yielding \scare{unfolded} vertices $\Runf$.
The potential energy obtained by summing over the original simulation cell, $U$, remains unchanged in this construction.
This allows to retain the small cutoffs needed for efficiency, while enabling \ad{} to compute the required derivatives.

With this construction, \cref{eq:J_general} can be rewritten as
\begin{align}
 \Junf &= \sum_{J \in \Runf} \frac{\partial \Bary}{\partial \R_J} \cdot \V_J \nonumber\\
  &\quad- \sum_{J \in \Runf}
    \left(\R_J \left(\dur{}{J}{} \cdot \V_J\right) \right) \, , \label{eq:J_unf}
\end{align}
introducing the energy barycenter
$\Bary = \sum_{I \in \Rsc}\nolimits \R_I U_I$, where the positions $\R_I$ are treated as pre-factors and not included in the partial derivative. The dot product is taken between denominator and velocity.
Writing the heat flux as the derivative of a vector $\Bary$ and a scalar $U$, as opposed to a high-dimensional Jacobian, ensures that these derivatives can be readily computed with \ad{}, incurring the same asymptotic computational cost as the calculation of $U$ and $\Bary$, which is proportional to $\magnitude{\Runf}$.
Since the number of additional positions is proportional only to the surface area of the simulation cell and the number of interactions $\interactions$, the overall asymptotic linear scaling is restored, with $\magnitude{\Runf} \propto N+N^{2/3}$ (see \cref{fig:heat_flux_timings}). 

To validate the approach, we benchmark the performance of a semi-local \mlp, in particular the SchNet~\cite{sktm2017q,sstm2018q} \mpnn architecture, for \gk calculations on zirconia (ZrO$_2$) and compare to results obtained with size-extrapolated \emph{ab initio} \gk~\cite{crs2017t}, as well as \gk with a local \mlp{}~\cite{vkjk2021q}, and experimental measurements~\cite{rwpm1998t,bflm2000t,mlld2004t}.

Training and validation data were generated using \emph{ab initio} \md in the $NpT$ ensemble, with four different trajectories heating up an initially tetragonal simulation cell with $96$ atoms to target temperatures \SI{750}{K}, \SI{1500}{K}, \SI{2250}{K} and \SI{3000}{K}.
In total, \num{10000} single-point calculations were performed using FHI-aims~\cite{FHI-aims} and FHI-vibes~\cite{FHI-vibes}, using the PBEsol~\cite{przb2008t} functional and otherwise following the computational approach of Ref.~\cite{crs2017t}.

On this data, we train a SchNet \mpnn{}, implemented in SchNetPack~\cite{sktm2018q}, with cutoff radius $\cutoff=\SI{5}{\angstrom}$. We choose an interaction depth $\interactions = 2$ leading to an effective cutoff of \SI{10}{\angstrom}. 
In line with recent findings by others~\cite{bkoc2022a}, we find this to be sufficient, as test set error does not significantly decrease for higher values of $\interactions$ or $\cutoff$.
Further details on the training procedure, choice of hyperparameters, and testing of the \mlp can be found in the \SM

We find that this simple approach yields a \mlp capable of describing the dynamics in monoclinic and tetragonal zirconia up to temperatures of approximately \SI{2000}{K}. In this temperature range, the anharmonic vibrational density of states matches that obtained from \dft. 
Beyond \SI{2000}{K}, the oxygen atoms become more mobile and different types of dynamical events are observed, in particular exchange-type oxygen diffusion. This behavior is also present in the training data in line with recent literature~\cite{tw2022t}, although slightly different diffusion events are observed given the smaller simulation cells and trajectory lengths. 
When diffusion increases at higher temperatures, the \mlp becomes unstable. This might be due to the limited amount of training data for these processes, especially for thermodynamic conditions close to the tetragonal-to-cubic phase transition. These observations suggest that an accurate description of defect formation is necessary to investigate zirconia above \SI{2000}{K}, which is beyond the scope of the current work.

\begin{figure}[t]
  \centering
  \includegraphics[scale=0.6]{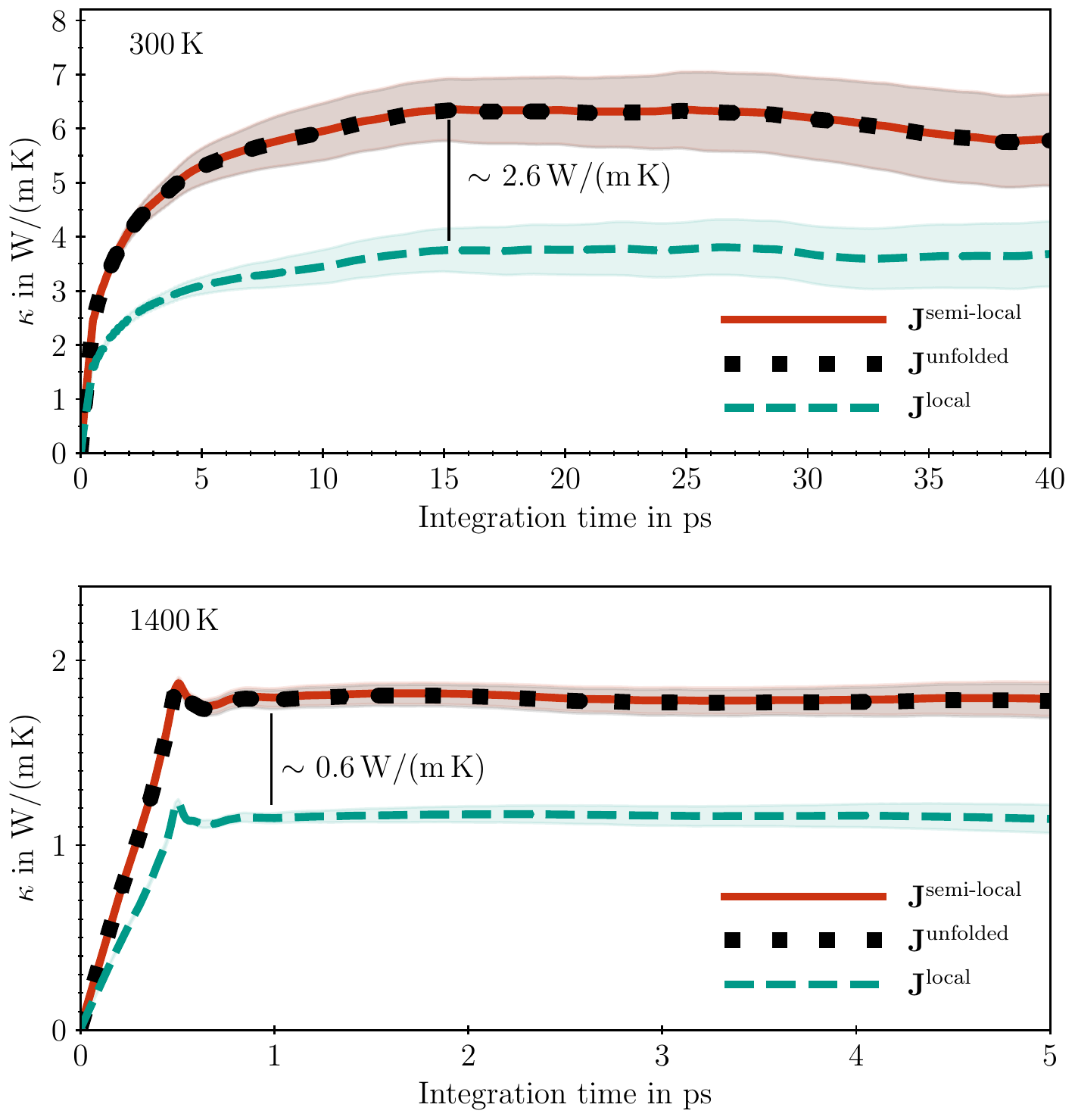}
  \caption{
  Comparison of the integral of the heat flux autocorrelation function for different formulations of the heat flux.
  The efficient re-formulation of the heat flux $\J^{\text{unfolded}}$ is equivalent to the full heat flux $\J^{\text{semi-local}}$, but not to
  $\J^{\text{local}}$, which neglects semi-local interactions.
  Results are given for an \mpnn{} with $\interactions{=\,}2$ and zirconia at \SI{300}{K} in the monoclinic phase (top) and \SI{1400}{K} in the tetragonal phase (bottom) for a simulation cell with 768 atoms.
  Shaded regions indicate standard error across eleven trajectories.
  }
  \label{fig:heat_flux_variants}
\end{figure}

\Cref{fig:heat_flux_variants} compares our efficient implementation with the full semi-local heat flux, as well as the purely local heat flux formulation. 
Due to the high computational cost of the unoptimized implementation, we use a small simulation cell with $N=\num{768}$ atoms, and rely on the noise reduction scheme introduced in Ref.~\cite{kcs2022t}.
The results confirm that our implementation $\Junf$ is equivalent to the semi-local heat flux $\Jmpnn$, while the local flux $\Jfan$ is not, underestimating the thermal conductivity by approximately \SI{40}{\percent} due to missing interactions beyond $\interactions = 1$.
A similar effect has been observed when formulations applicable to pairwise additive potentials are used for many-body force fields~\cite{bbw2019t,smko2019t}.

\begin{figure}[t]
  \centering
  \includegraphics[scale=0.6]{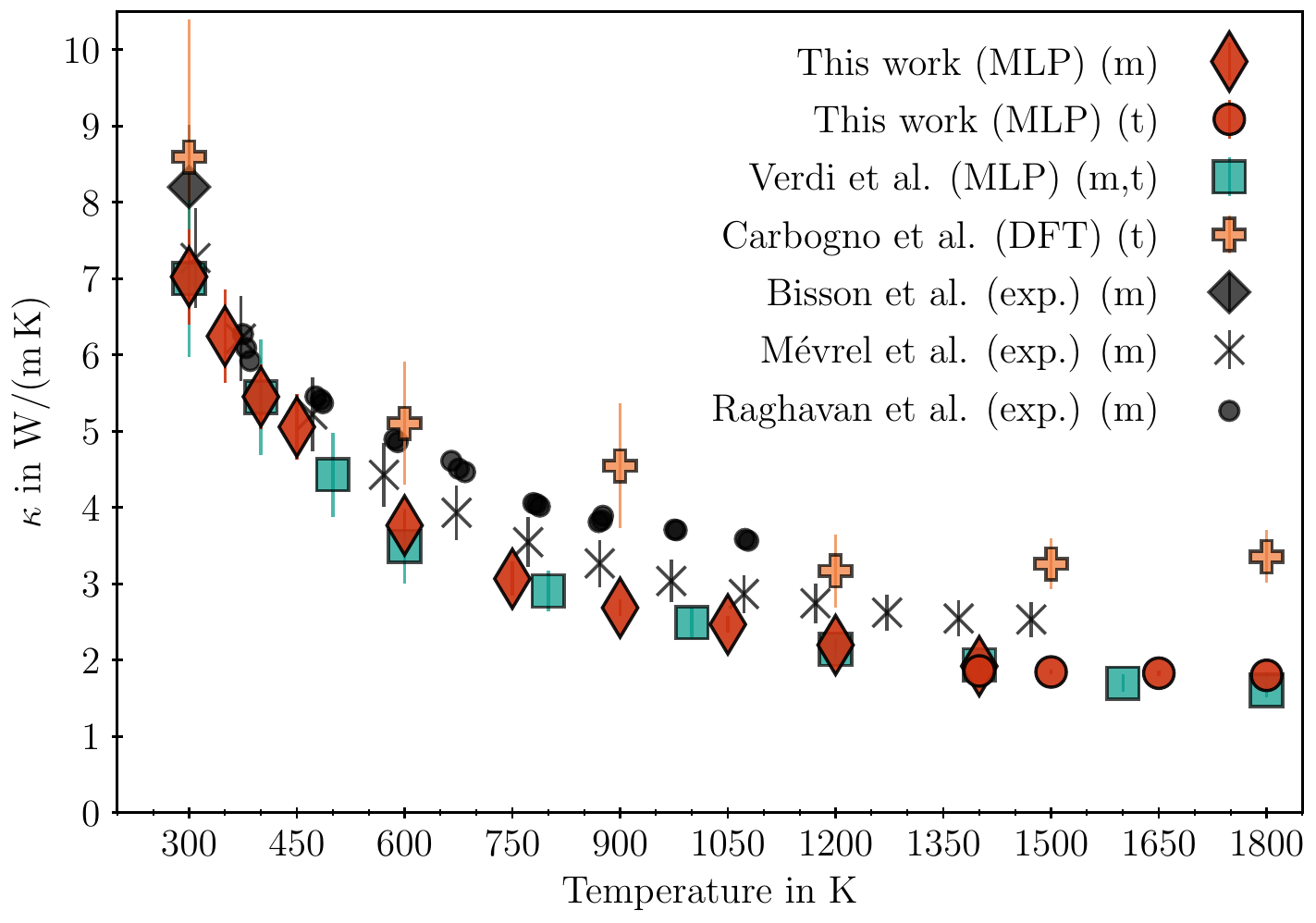}
  \caption{Thermal conductivity across temperatures computed with an \mpnn{} using $\interactions{=\,}2$ message-passing steps
  and experimentally determined lattice parameters~\cite{ps1969t,kh1998t},
  compared with
  another MLP without extrapolation~\cite{vkjk2021q},
  size-extrapolated \emph{ab initio} \gk~\cite{crs2017t},
  and
  experimental measurements~\cite{rwpm1998t,bflm2000t,mlld2004t}.
  Error bars are shown as given in the respective publications, the ones in the present work reflect the standard error across eleven trajectories. Letters \scare{t} and \scare{m} indicate results for the tetragonal and monoclinic phase, respectively.
  }
  \label{fig:kappa_vs_temperature}
\end{figure}

Enabled by computationally efficient access to $\J$ for semi-local \mlps{}, we then predict the thermal conductivity of zirconia across temperatures. Since the focus of the present work is the heat flux, we do not treat the thermodynamics of zirconia with the \mlp{}, but use experimentally determined lattice parameters~\cite{ps1969t,kh1998t} to account for lattice expansion. At \SI{1400}{K}, both phases are investigated, as the monoclinic phase is sufficiently stable during the course of the simulations, which consist of eleven trajectories of \SI{1}{ns} each, with a $N=\num{1500}$ simulation cell. These settings yield fully size- and time-converged results (see \SM{} for details).

The results presented in \cref{fig:kappa_vs_temperature} are in good agreement with both experimental measurements in the monoclinic phase, and theoretical \mlp{} predictions in the monoclinic and tetragonal phases.
As this work uses similar lattice parameters and the same exchange-correlation functional as the work by Verdi~\etal{}~\cite{vkjk2021q}, the observed close agreement is to be expected.
Remaining differences between the \mlp results may be due to larger simulation cells used in the present work, enabled by the favorable scaling of computational cost due to the efficient heat flux implementation, and the semi-local nature of the employed \mpnn.
Compared to experiment, both \mlps are found to systematically underestimate $\kappa$ by approximately \SIrange{10}{20}{\percent}, which may be related to the intrinsic approximation of a finite-range \mlp{}, or the underlying density functional approximation.

Larger differences are observed with the size-extrapolated \emph{ab initio} \gk results reported by Carbogno~\etal~\cite{crs2017t}, which, however, were computed for the tetragonal phase at all temperatures. Additionally, due to the high computational cost of first-principles calculations, only three trajectories of \SI{60}{ps} each were used, which is reflected in the larger statistical error. 

We conclude that the adapted GK approach for semi-local \mlps{} introduced in this work can successfully and efficiently predict the thermal conductivity of zirconia across temperatures, using \num{10000} first-principles calculations in total. Despite a moderate system size of 96 atoms for training, fully size-converged results were obtained without requiring additional extrapolation schemes.

In summary, we have demonstrated the feasibility of applying \ad{}-based semi-local \mlps{} to the prediction of thermal conductivities with the \gk method.
For this, we investigated the impact of semi-local interactions on the heat flux, and derived an adapted heat flux $\Junf$ that can be efficiently implemented via \ad.
This heat flux has asymptotically linear runtime and requires no further restrictions on the form of the potential. Its formulation is independent of the body-order of the potential energy function, making no distinction between pair, angle-dependent, or many-body potentials. As it relies on explicitly constructing an extended simulation cell, it is applicable to semi-local \mlps{} with moderate effective interaction ranges.

% \vspace{2\baselineskip}
\section*{Data and Code Availability}
\noindent
Data and code required to reproduce all figures can be found at \href{https://doi.org/10.5281/zenodo.7767432}{doi:10.5281/zenodo.7767432}. First-principles calculations for the training data are additionally available on the NOMAD repository at \href{https://doi.org/10.17172/NOMAD/2023.03.24-2}{doi:10.17172/NOMAD/2023.03.24-2}. 
Further information and software can be found in the \SM{} and at \href{https://marcel.science/gknet}{https://marcel.science/gknet}.

% \vspace{2\baselineskip}
\section*{Acknowledgements}
\noindent
This work was supported by the TEC1p Project (ERC Horizon 2020 No. 740233).
M.F.L. gratefully acknowledges financial support by the German Ministry for Education and Research BIFOLD program (refs. 01IS18025A and 01IS18037A).
F.K. acknowledges support from the Swedish Research Council (VR) program 2020-04630, and the Swedish e-Science Research Centre (SeRC).
M.R. acknowledges funding from the European Union's Horizon 2020 research and innovation program under Grant Agreement 952165.
Part of this research was performed while M.F.L. and M.R. were visiting the Institute for Pure and Applied Mathematics (IPAM), which is supported by the National Science Foundation (Grant No. DMS-1925919).
M.F.L. would like to thank
Profs.~Klaus-Robert Müller and Davide Donadio,
as well as
Carla Verdi, Fabian Nagel, and Adam Norris for constructive discussions and support.

\bibliography{babel_short}

\end{document}

\title{Supplementary Material:\\Heat flux for semi-local machine-learning potentials}

\author{Marcel F. Langer}
\email[Corresponding author: ]{mail@marcel.science}
\affiliation{Machine Learning Group, Technische Universit{\"a}t Berlin, 10587 Berlin, Germany}
\affiliation{BIFOLD -- Berlin Institute for the Foundations of Learning and Data, Berlin, Germany}
\nomadaff

\author{Florian Knoop}
\email[Corresponding author: ]{florian.knoop@liu.se}
\affiliation{Theoretical Physics Division, Department of Physics, Chemistry and Biology (IFM), Linköping University, SE-581 83 Linköping, Sweden}
\nomadaff

\author{Christian Carbogno}
\nomadaff

\author{Matthias Scheffler}
\nomadaff

\author{Matthias Rupp}
% \email[Corresponding author: ]{rupp@mrupp.info}
\nomadaff
\affiliation{Department of Computer and Information Science, University of Konstanz, 78464 Konstanz, Germany}
\affiliation{Materials Research and Technology Department, Luxembourg Institute of Science and Technology (LIST), Belvaux, Luxembourg}

\maketitle

\section{Automatic Differentiation}

For practical simulations based on machine-learning potentials, it is not only necessary to predict the total energy~$U$, but also its derivatives with respect to atomic positions $\R_I$,~e.g.,~the forces and the stress tensor~\cite{kcbs2015t}.
Since it is impractical to manually implement the required derivatives for deep neural networks or other complex architectures,  automatic differentiation (\ad{}) is employed.

\ad{} is a technique to automatically obtain derivatives of functions implemented as computer programs~\cite{griewank2008,bprs2017m} by reducing them to a directed acyclic graph of elementary operations, evaluating the local derivatives for each operation, and then applying the chain rule by traversing the graph: Traversing the graph from the output backwards yields reverse-mode \ad{}. Traversing from the inputs forwards yields forward-mode \ad{}.

\ad{} can efficiently obtain vector products of the Jacobian of a function over either the input dimensions (forward mode) or the output dimensions (reverse mode), with the same asymptotic runtime as computing the original function. Computing the explicit Jacobian, on the other hand, requires repeated evaluations of such products, incurring linear additional computational cost in either the input or output dimension.

Therefore, if the energy prediction scales linearly with system size, and only derivatives of the scalar total potential energy $U$ are required, the forces and stress can be predicted in a linearly-scaling fashion as well. In contrast, explicitly computing all derivatives of the potential energy \emph{contributions}, $\indur{I}{J}{}$, would incur quadratic computational cost, which introduces difficulties when implementing the \gk{} method.

Since \ad{} relies on tracing the underlying computation, it can only produce derivatives with respect to quantities that were explicitly used in computing a result; for instance, derivatives with respect to pair vectors $\R_{IJ}$ cannot be computed when $J \notin \nbh{I}$, i.\,e., for atoms that are not in the neighborhoood of $I$, since those atom-pair vectors only appear implicitly through message passing, which is discussed in the next section.

\section{Semi-Local Machine-Learning Potentials}
\label{sub:si-mlp}

\begin{figure*}
    % \begin{tikzpicture}[
    %     x = 6.5cm,
    %     y = 3.0cm,
    %     >=stealth,
    %     ]
    %     \node (a) at (0,1) {\input{img/fig_mpnn_sketch_1}};
    %     \node (b) at (1,1) {\input{img/fig_mpnn_sketch_2}};
    %     \node (c) at (2,1) {\input{img/fig_mpnn_sketch_3}};
    %     \node (d) at (0,0) {\input{img/fig_mpnn_sketch_4}};
    %     \node (e) at (1,0) {\input{img/fig_mpnn_sketch_5}};
    %     \node (f) at (2,0) {\input{img/fig_mpnn_sketch_6}};
    %     \draw [->] (a) -- (b) ;
    %     \draw [->] (b) -- (c) ;
    %     \draw [->] (c) -- (2,0.5) -- (0,0.5) -- (d) ;
    %     \draw [->] (d) -- (e) ;
    %     \draw [->] (e) -- (f) ;
    % \end{tikzpicture}
    \includegraphics[scale=1.0]{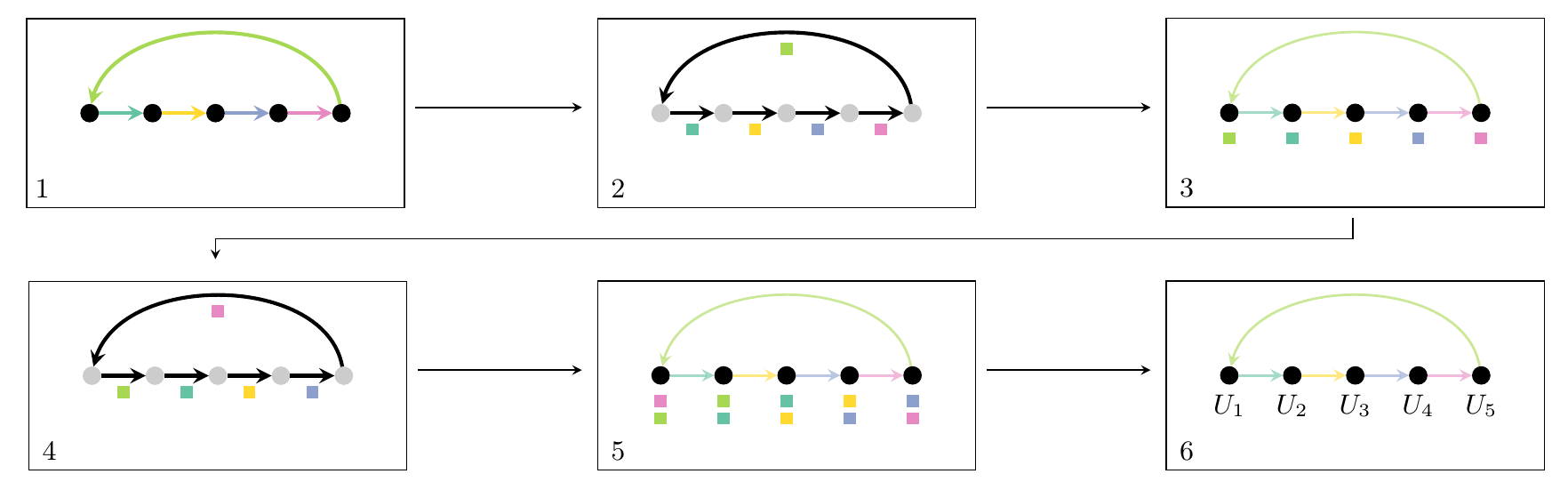}
    \caption{Sketch of neural message passing with $\interactions=2$. Connections are only considered in one direction for simplicity. 
    \label{fig:si-mpnn_sketch}
    \textit{(1)} Initialization: Each edge symbolises an interatomic distance and is associated with a color. Each node is an atom with an empty initial state.
    \textit{(2+3)} Message-passing: Each node is updated based on the (empty) neighboring state, and its incoming edge. Afterwards, each node depends on the incoming edge.
    \textit{(4+5)} Another message-passing iteration. Now, every state depends on next-to-nearest neighbors.
    \textit{(6)} Readout: $U_I$ are predicted based on the final states.
    }
\end{figure*}

In this work, we consider machine-learning potentials (\mlps{}) with semi-local structure. In many cases, \mlps{} of this type are implemented as message-passing neural networks (\mpnns{})~\cite{gsvd2017q}, although constructions based on other forms of regression are also possible~\cite{blcd2022a}.

This type of model aims to combine the computational efficiency of local \mlps{}, in which potential energy contributions are dependent on atoms within a cutoff $\cutoff$, typically on the order of \SIrange{5}{10}{\angstrom}, with the ability to model correlations extending beyond these cutoffs. This is achieved by allowing $\interactions$ interactions between neighboring environments, building up longer-range correlations iteratively, without requiring the explicit construction of the extended neighborhoods, at the expense of an information bottleneck introduced by the restricted amount of information that can be communicated. One can alternatively view semi-local models as introducing a sparsification of interactions within $\effcutoff = \interactions \cutoff$. 

In order to clarify the construction of such models in practice, we now briefly introduce \mpnns{}.
The (periodic) system at hand is represented by a graph $\graph = (\vertices, \edges)$, where the vertices $\vertices$ correspond to the atoms present in the simulation cell connected by edges $\edges$. 
Two vertices $I$ and $J$ are connected with an edge if $|\R_{IJ}| \leq r_{\text{c}}$, where $\cutoff$ is a user-defined cutoff radius. Since, by definition, $I$ and $J$ are atoms in the simulation cell, $\R_{IJ}$ must be subjected to the minimum image convention (\mic{}) to account for periodicity,
 % defined as 
\begin{equation}
     \Rm_{IJ} = \min_{\offset \in \mathbb{Z}^3} \magnitude{\R_J + \sum_a n_a \basis_a - \R_I} \,,
\end{equation}
in which $\basis_{a=1,2,3}$ denote the basis vectors spanning the simulation cell.
Edges $\edges$ are then labelled with the \mic{} atom-pair vectors.
All vertices $J$ connected with an edge to $I$ define the local atomic neighborhood $\nbh{I}$.
$\graph$ is therefore \emph{boundary invariant}, i.e., independent of the particular choice of unit cell; translational invariance is ensured by using relative positions.

\mpnn{} potentials operate on $\graph$ in three stages, illustrated in \cref{fig:si-mpnn_sketch}: initialization, message-passing, and readout. Each message-passing step is labeled by $t = 0, \ldots, \interactions$. During initialization ($t=0$), every atom $I$ is assigned a state vector $\state{I}{t=0}$, based on chemical species. 
% \flocomment{I suggest to use and stick to "receiver atom" instead of "central, receiving atom"} -> OK
During the next stage, these states are \emph{updated} with \emph{messages} $\msg{I}{t}$ that can depend on the states of both the receiver atom $I$ and its neighbors $J \in \nbh{I}$, as well as the edges $\R_{IJ}$ connecting them:
\begin{align}
    \msg{I}{t+1} &= \sum_{J \in \nbh{I}} \msgf{t}(\state{I}{t}, \state{J}{t}, \edge{I}{J}) \\
    \state{I}{t+1} &= \updatef{t}(\state{I}{t}, \msg{I}{t+1}) \, .
\end{align}
Here, the message functions $\msgf{t}$ and update functions $\updatef{t}$ are differentiable functions implemented as neural networks. They are learned during training, and can differ over message-passing iterations. 
The message functions $\msgf{t}$ model pairwise interactions. Their aggregation into the total message $\msg{I}{t+1}$ via a sum ensures permutational invariance.
The update functions $\updatef{t}$ describe how the interactions with all neighbors influence the new state of the receiver atom. Over multiple iterations, information is \emph{propagated} through the graph.

We emphasize that this construction incorporates periodicity by messages wrapping around boundaries back into the simulation cell. 
Since the atom-pair vectors within $\nbh{I}$ are identical whether $I$ is in the simulation cell or a replica, this is equivalent to messages propagating into neighboring replicas. However, as we will see in the following section, this implicit treatment of periodicity introduces difficulty with the heat flux, which is resolved by adapting $\graph$.

After $\interactions$ message-passing steps, the final stage is reached where a readout function $R$ predicts the atomic potential energy contributions $U_I$. This atom-independent readout function is learned during training and acts on each vector $\state{I}{\interactions}$ representing the final state of atom~$I$. The total potential energy $U$ is then simply given by
\begin{align}
  U = \sum_{I \in \vertices} U_I
  \qquad \text{with} \qquad
  U_I = R(\state{I}{\interactions})~.
%   U_I &= R(\state{I}{\interactions})~, & U &= \sum_{I \in \vertices} U_I \, .
  \label{eq:mpnn_ui} 
\end{align}
For $\interactions>1$, the energy contribution $U_I$ is a function not only of $\{ \edge{I}{J} | J \in \nbh{I} \}$, but of every edge within $\interactions$ hops on the graph: Information propagates beyond local neighborhoods of atoms, allowing for implicit long-range interactions. This distinguishes \mpnn{} potentials from other MLPs, which typically act only on local features. In the present framework, such simpler, strictly local models are included when setting $\interactions = 1$.

We also note that different strategies can be employed to deal with rotational invariance. Early \mpnns{} such as SchNet~\cite{sktm2017q,sstm2018q} only use interatomic distances as inputs, discarding angular information. 
Recent equivariant \mpnns{} \cite{tskr2018q,ahk2019q,bmsk2022q,ucsm2021q,sug2021a} can make use of atom-pair vectors and can ensure that $\state{I}{t}$ and $\msg{I}{t}$ transform appropriately under rotations. 
Message functions that take angular or higher-order information into account are also possible~\cite{kgg2020q,bkoc2022a,bbkc2022a,fum2022a}. In this work, we use an invariant \mpnn{}. The presented derivations and conclusions are however fully general and applicable to equivariant \mpnns{}.

\section{Heat Flux Formulations}
\label{sub:si-hf}

\subsection{Derivation of Hardy Formula}

Due to the difficulties discussed in the main text, deriving and implementing the heat flux for particular types of \pes has been the focus of much previous work~\cite{c2006t,at2011t,tno2008t,fpdh2015t,mub2015t,crs2017t,smko2019t,bbw2019t}.
This work aims to provide a unified perspective for the case of semi-local \mlps implemented using \ad.

As a starting point for this task, we re-derive the classical equivalent of the heat flux introduced by Hardy~\cite{h1963t}, combining existing approaches with the aim of clarifying the introduction of periodic boundary conditions. Often, the Hardy formula is given without explicitly specifying the range of the involved sums, introducing ambiguity in periodic boundary conditions: Sums can run either over the simulation cell, or the bulk, and describe either \scare{checkerboard} or \scare{toroidal} boundary conditions, an ambiguity highlighted by Erpenbeck~\cite{e1995t} in the context of \gk{} calculations of shear coefficients. In the context of \ad{}, which relies explicitly on the forward computation of $U$, the precise construction of a given potential gains central importance, and we therefore take care to be as explicit as possible.

In this derivation, we rely on the approach by Hardy~\cite{h1982t} to provide a solution to the continuity equation in terms of bond functions,\footnote{A similar, but more rigorous machinery was developed by Noll~\cite{n1955t}, extending the work of Irving and Kirkwood~\cite{ik1950t}, but in the interest of brevity, we do not adopt it here.} and follow the ideas of Ito and Nakamura~\cite{in2009t} to avoid the assumption of a pair potential. After solving the continuity equation for a finite system of $N$ particles, we adopt periodic boundary conditions, recovering the Hardy formulation of the heat flux~\cite{h1963t,fpdh2015t}.

The heat flux required for the \gk{} method is defined as the spatial integral of a current density
\begin{equation}
    \J(t) = \integral{V}^3r \, \densj(\vR, t) \, ,
\end{equation}
which arises from a continuity equation for the energy density
\begin{equation}
    \dot e(\vR, t) + \boldsymbol{\nabla} \cdot \densj(\vR, t) = 0 \, , \label{eq:hf_ceqn}
\end{equation}
which in turn is defined as a scalar field that integrates to the total energy of the system
\begin{equation}
    E(t) = \integral{V}^3r \, e(\vR, t) \, .
\end{equation}

We are studying the system under consideration with \md{}, treating it as a collection of $N$ point particles contained in the volume $V$, whose state is fully characterised by a set of phase-space coordinates
$\fullphasp = \curlyset{\phasp{I}{}}{I=1 \ldots N} = \curlyset{(\R_I, \momentum_I)}{I=1 \ldots N}$, with atomic momenta $\momentum_I$.
These coordinates evolve in time according to Newton's equations of motion
\begin{equation}
    \F_I = - \partial U(\fullphasp) / \partial \R_I = \dot \momentum_I = m_I \ddot \R_I \, , \label{eq:eom}
\end{equation}
where $U$ is the potential energy, which we take to be composed of atomic contributions; $U = \sum_{I=1}^N\nolimits U_I$.
This work considers potentials where this decomposition is available by design, as they approximate the \bo \pes in terms of many-body functions $U_I$ of the atomic coordinates.

In \dft, on the other hand, the potential energy part of the Hamiltonian has a strictly pairwise structure; many-body interactions then emerge \emph{implicitly} from the electronic density, instead of being explicitly included in the energy expression.
Carbogno~\etal~\cite{crs2017t} exploit this pairwise structure, which extends to the Hellmann-Feynman forces, to obtain pairwise derivatives of the potential. They then obtain an exact equivalent of $\Jpot$, which purely relies on derivatives, and discard $\Jconv$.
The meaning and naming of these terms is further discussed at the end of the present section.
We additionally note that an energy density for \dft can be defined~\cite{cm1992t} and a particular gauge can be chosen to obtain a heat flux~\cite{mub2015t}.

We now proceed with a derivation for potentials where atomic potential energies $U_I$ are explicitly available, and for the sake of generality assume that each $U_I$ can depend on the position of every atom $J$, such that $U_I = U_I(\{\R_J\})$. We will later specialize to particular types of potentials, restricting the dependence of $U_I$ to finite neighborhoods.

With this, time-dependence appears only implicitly through the time-dependent positions and velocities, allowing us to recast the total time derivative in terms of partial derivatives with respect to positions. We suppress time-dependence in the following derivation. Additionally, all sums will run from $1$ to $N$ until the spatial integral over $\densj(\vR)$ is executed.

Following Hardy~\cite{h1982t}, we spatially localize the atomic total energy, given by the sum of potential energy $U_I$ and kinetic energy $T_I$, $E_I = U_I + T_I$ with a function $\localf{\R_I - \vR}$, which is peaked at $\R_I$, decays to $0$ as $|\R_I - \vR|$ increases, and is normalized to $1$. 

We also define a corresponding bond function, which localizes contributions along a line segment:
\begin{equation}
    \bondf{IJ}{\vR} = \int_0^1 \, \text{d} \lambda\, \localf{\lambda \R_I + (1-\lambda) \R_J - \vR} \, .
\end{equation}
One can then show that
\begin{align}
    \R_{IJ} \cdot \grad_{\vR} \bondf{IJ}{\vR} = \localf{\R_I - \vR} - \localf{\R_J - \vR} \, .
\end{align}
We then make an ansatz for $\dense(\vR)$:
\begin{align}
    \dense(\vR) = \sum_I \localf{\R_I - \vR} E_I \, .
    % \dense(\vR) &= \sum_I \localf{\R_I - \vR} \left(U_I + T_I\right) \\
    % &= \sum_I \localf{\R_I - \vR} E_I \, ,
\end{align}
% with kinetic energy $T_I$ and atomic potential energy $U_I$.
The time-derivative of $\dense(\vR)$ is
\begin{align}
    \totaldiff{t} \dense(\vR, t) &= \sum_I \left(\totaldiff{t} E_I\right) \localf{\R_I - \vR} \nonumber \\
    &\quad+ \sum_I E_I \left(\totaldiff{t} \localf{\R_I - \vR} \right) \, .
\end{align}
The first term can be tackled by again splitting $E_I = U_I + T_I$ and resolving the time derivative using \cref{eq:eom}: 
\begin{align}
    \totaldiff{t} E_I &= \totaldiff{t} U_I + \totaldiff{t} T_I \\
    &= \sum_j \dur{I}{J}{} \cdot \V_J + \sum_I F_i \cdot \V_I \\
    &= \sum_j \dur{I}{J}{} \cdot \V_J - \sum_I \dur{}{I}{} \cdot \V_I \, .
\end{align}
Then, we re-arrange it in a pair-wise form
\begin{align}
    &\sum_I \left(\totaldiff{t} E_I\right) \localf{\R_I - \vR}  \\
    &= \sum_I \left(\sum_j \dur{I}{J}{} \cdot \V_J \right) \localf{\R_I - \vR} \\
    &\quad- \sum_I \left(\dur{}{I}{} \cdot \V_I \right) \localf{\R_I - \vR} \\
    &= \sum_{IJ} \left(\dur{I}{J}{} \cdot \V_J\right)  (\localf{\R_I - \vR} \nonumber \\
    &\qquad\qquad\qquad\qquad\qquad- \localf{\R_J - \vR}) \, ,
\end{align}
and finally apply the identity from above to obtain
\begin{equation}
    = \grad_{\vR} \cdot \left[\sum_{IJ} \R_{IJ} \left(\dur{I}{J}{} \cdot \V_J\right) \bondf{IJ}{\vR} \right] \, .
\end{equation}

The second term is easily resolved by the chain rule
\begin{align}
    &\sum_I E_I \left(\totaldiff{t} \localf{\R_I - \vR} \right) \\
    &= -\sum_I E_I \Big( \grad_{\vR} \localf{\R_I - \vR} \cdot \V_I \Big) \\
    &= -\grad_{\vR} \cdot \Big( \sum_I E_I \localf{\R_I - \vR} \V_I \Big) \, .
\end{align}
Comparing with the continuity equation, we find
\begin{align}
    \densj(\vR)
    & = \sum_{IJ} \R_{JI} \left(\dur{I}{J}{} \cdot \V_J\right) \bondf{IJ}{\vR} \nonumber\\
    &\quad + \sum_I E_I \V_I\, \localf{\R_I - \vR} \, , \label{eq:hf_densj}
\end{align}
a direct solution to the continuity equation.

This result can also be justified from a physical perspective. Any energy change at atom $I$ must be balanced out by corresponding changes in the atoms $J$ it interacts with, with an energy current flowing between them.\footnote{This does not mean that pairwise fluxes are always zero -- this is only the case if $\indur{I}{J}{} = \indur{J}{I}{}$, which only applies to pair potentials, as pointed out by Fan~\etal~\cite{fpdh2015t}.} If we can divide up the change in $E_I$ into contributions that can be attributed to different $J$, and identify the corresponding terms in $E_J$, we know the magnitude of the current between $I$ and $J$. It is a natural assumption that the current flows directly between them, parallel to $\R_{IJ}$. The first term is one way to construct a vector field accordingly.
The second term describes an alternative process to change local distribution of energy: Rather than energy flowing between atoms, the energy assigned to a given atom is \scare{dragged along} by the atom moving. Together, these two processes describe how the energy density can change, and therefore yield the energy current density, solving the continuity equation.

We are now in a position to compute the macroscopic heat flux $\J$. However, we must first define the integration domain: Since the thermodynamic limit is inaccessible due to finite computational resources, practical simulations make use of periodic boundary conditions, replicating a simulation cell with $N$ atoms periodically in space. This simulation cell must be large enough to capture the relevant dynamic processes for thermal transport. We then formally compute $\densj$ for the resulting bulk, defining $\Rsc$ as the atoms in the simulation cell and $\Rall$ as the entire bulk.
Extending the so-far unspecified sum over $IJ$ explicitly over the bulk yields
\begin{align}
    \densj(\vR)
    &= \sum_{\substack{I \in \Rall \\ J \in \Rall}} \R_{JI} \left(\dur{I}{J}{} \cdot \V_J\right) \bondf{IJ}{\vR} \nonumber \\
    &\quad+ \sum_I E_I \V_I \localf{\R_I - \vR} \, .
\end{align}
We now integrate over the simulation cell, obtaining
\begin{align}
    \J &= \sum_{\substack{I \in \Rsc \\ J \in \Rall}} \left(\R_{JI} \left(\dur{I}{J}{} \cdot \V_J\right) \right) \nonumber \\
    &\quad+ \sum_{I \in \Rsc} E_I \V_I \, ,\label{eq:si-hf_general}
\end{align}
where we have used the fact that all contributions where $I \notin \Rsc$ can be computed equivalently for the replica of $I$ in the simulation cell. This is a general form of the heat flux, the classical equivalent of the formulation derived by Hardy for periodic quantum systems~\cite{h1963t,fpdh2015t}. We note that the minimum image convention does not yet appear.

\Cref{eq:si-hf_general} is typically split into two terms: The first one is identified as the \scare{potential} heat flux $\Jpot$, describing heat flux due to energy exchange between atoms, and the second one as \scare{convective} term $\Jconv$, accounting for changes in energy density due to atomic movements directly.
Nomenclature changes across the literature. For instance, Fan \etal~\cite{fpdh2015t} refer to $\Jconv$ as \scare{kinetic} term, while Carbogno \etal~\cite{crs2017t} refer to $\Jpot$ as \scare{conductive} or \scare{virial} term. We do not adopt this terminology as $\Jpot$ is not always composed of virials, i.e., contributions to the stress: In general semi-local \mlps, the derivative $\indur{I}{J}{}$ does not directly resolve into atom-pair vectors originating from $I$.
In solids, $\Jpot$ is also not precisely the heat flux that determines thermal conductivity.

\subsection{Heat Flux in Solids}

\begin{figure}
  \centering
  \includegraphics[scale=0.6]{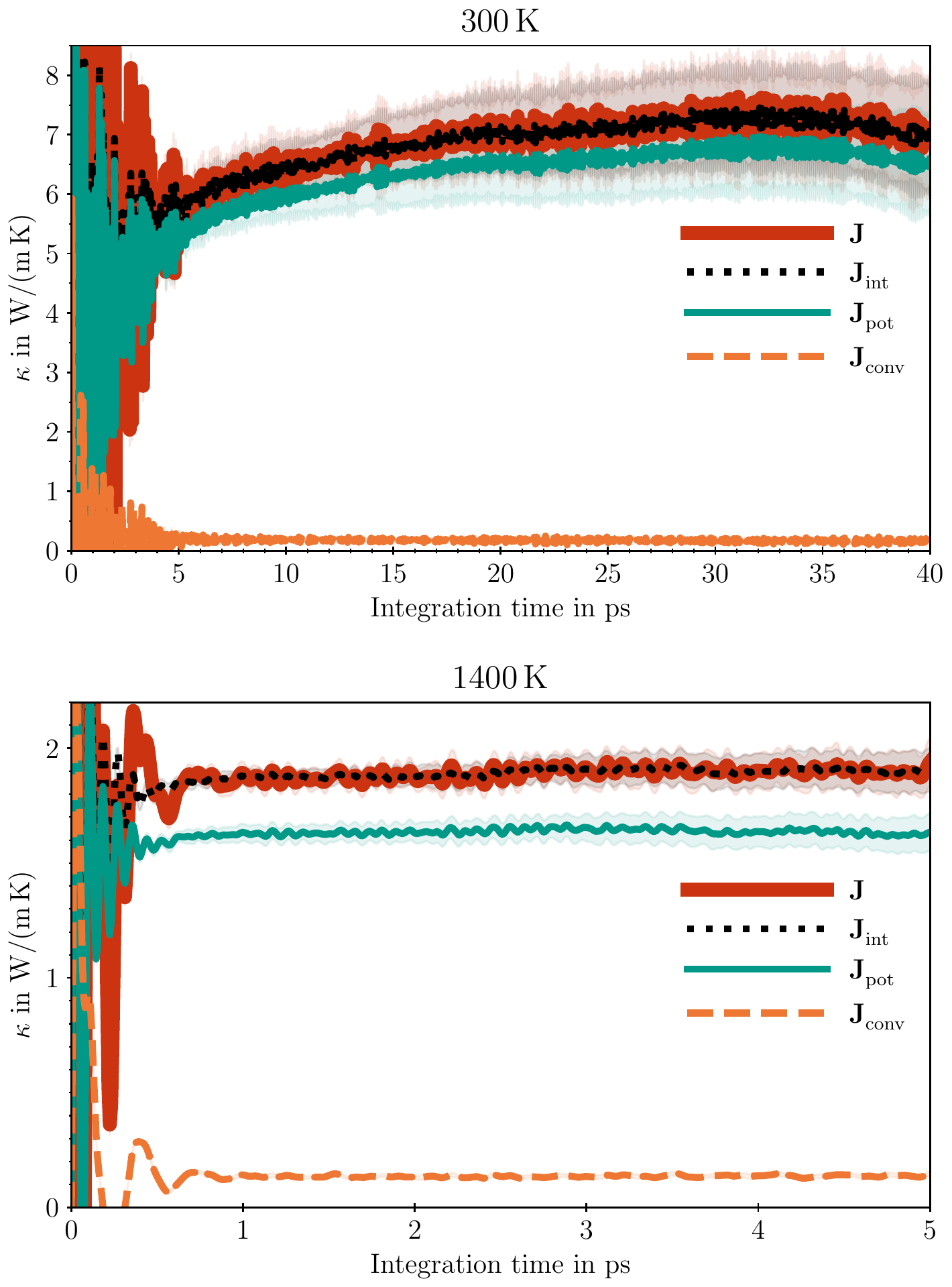}
  \caption{Comparison of the integral of the heat flux autocorrelation function for $\J$, $\Jint$, $\Jpot$ and $\Jconv$, at \SI{300}{K} in the monoclinic phase (top) and \SI{1400}{K} in the tetragonal phase (bottom), using \scare{production} settings ($N{=}1500$, $t{=}\SI{1}{ns})$.
  No noise reduction is applied in this case.
  Shaded areas indicate standard error.
  }
  \label{fig:si-heat_flux_convective}
\end{figure}

In solids, atomic positions are bounded over time. In other words, atomic positions $\R_I(t)$ can be split into a fixed \emph{reference} position $\Rr_I$ and a time-dependent displacement from that position $\U_I(t)$:
\begin{equation}
    \R_{I}(t) \defas \Rr_{I} + \U_I(t) \quad I \in \Rall \\
\end{equation}
If we choose $\Rr_I$ such that it contains the information about which replica $I$ belongs to,\footnote{For instance by choosing positions in the pristine supercell, or $\Rr_I = \R(t=0)$, or average positions over the simulation run.} we obtain:
\begin{align}
    \R_{I\offset}(t) &= \Rr_{I\offset} + \U_I(t) \\
    \Rightarrow \dot \R_{I\offset}(t) &= \V_I(t) \, ,
\end{align}
where we introduce $\offset$ as a an offset vector, indicating which integer multiples of lattice vectors are added to yield a given replica.
In other words, the displacements and velocities are shared between all replicas of $I$. Substituting into \cref{eq:si-hf_general}, we obtain:
\begin{align}
    \J &= 
        \sum_{\substack{I \in \Rsc \\ J \in \Rall}} \left(\Rr_{JI} \left(\dur{I}{J}{} \cdot \V_J\right) \right)
        + \totaldiff{t} \sum_{I \in \Rsc} \U_I E_I \label{eq:hf_jfull} \\
        &\defdas \Jint + \Jdiff \, . 
\end{align}
This defines a new split for $\J$, into an \scare{interaction} term $\Jint$, which arises due to the exchange of energy between atoms at sites $I$ and $J$, and a remainder term due to the displacements, $\Jdiff$.
If $\U_I(t)$ are bounded over time, then $\Jdiff$ is also bounded, and is therefore a non-diffusive heat flux, which does not contribute to $\tk$~\cite{lmh1986t,mub2015t,ibdb2019t}.

As we consider a solid system in this work, only $\Jint$ is reported throughout, as it allows for a more straightforward implementation and noise removal. \Cref{fig:si-heat_flux_convective} compares $\J$ and $\Jint$, showing that they are indeed equivalent for zirconia at the given temperatures. We note that $\Jconv$ does not disappear in isolation, and that $\Jpot$ is generally \emph{not} equivalent to $\Jint$.

\subsection{Local Heat Flux}

Let us briefly consider the heat flux for general local potentials where
\begin{equation}
  U_I = U_I(\nbh{I}) \, ,
\end{equation}
with $\nbh{I}$ containing positions, \emph{including replicas}, up to a finite cutoff radius $\cutoff$.
Then
\begin{align}
    &\J = \sum_{\substack{I \in \Rsc \\ J \in \nbh{I}}} \left(\R_{JI} \left(\dur{I}{J}{} \cdot \V_J\right) \right) 
        + \sum_{I \in \Rsc} E_I \V_I \, .
\end{align}
Since potentials are expected to be translationally invariant, the dependence on $\R_J$ can only enter through atom-pair vectors, and we therefore obtain
\begin{align}
  &\J = \sum_{\substack{I \in \Rsc \\ J \in \nbh{I}}} \left(\R_{JI} \left(\dur{I}{I}{J} \cdot \V_J\right) \right) 
        + \sum_{I \in \Rsc} E_I \V_I \, ,
\end{align}
the form derived by Fan~\etal{}~\cite{fpdh2015t}. If we further note that no interactions between environments occur, then every atom-pair vector can be uniquely assigned to one $U_I$, and we can equivalently write
\begin{align}
  &\J = \sum_{\substack{I \in \Rsc \\ J \in \nbh{I}}} \left(\R_{JI} \left(\dur{}{I}{J} \cdot \V_J\right) \right) 
        + \sum_{I \in \Rsc} E_I \V_I \, .
\end{align}
A similar form was derived by Carbogno~\etal{}~\cite{crs2017t} for \dft{}, using the pairwise nature of the derivatives of the Hamiltonian. However, in \dft, electronic degrees of freedom enter $U$.

In the present work, we consider local potentials as the $\interactions=1$ special case of semi-local models as introduced in \cref{sub:si-mlp}. 
In this setting, replicas are not explicitly constructed. Instead, the \mic{} is used on the atom-pair vectors providing the edges in the graph. The heat flux for the $\interactions=1$ case is therefore
\begin{align}
  \J &= \sum_{IJ \in \edges} \left(\R_{JI} \left(\dur{I}{I}{J} \cdot \V_J\right) \right) 
        + \sum_{I \in \Rsc} E_I \V_I \\
    &= \sum_{IJ \in \edges} \left(\R_{JI} \left(\dur{}{I}{J} \cdot \V_J\right) \right) 
    + \sum_{I \in \Rsc} E_I \V_I \, .
\end{align}
This form can be efficiently implemented with \ad{}, as it only requires derivatives of the scalar total energy, which can be obtained in one backwards pass.

\subsection{Semi-Local Heat Flux}

To treat semi-local \mlps{}, we first note that by construction, \cref{sub:si-mlp}, interactions beyond the simulation cell are re-mapped into the cell and treated \emph{implicitly}. \Cref{eq:si-hf_general} cannot be immediately applied, since the replicas in $\Rall$ are not explicitly constructed, and are therefore inaccessible to \ad.
The approach used for local potentials, where we used the fact that each edge can be uniquely assigned to a given $U_I$ is also not applicable, as $\interactions > 1$ correlates neighboring environments.

A general solution to this difficulty is provided by the \scare{unfolded} heat flux presented in the following section, which changes the construction of semi-local \mlps{} to allow this attribution.
A preliminary solution that respects the standard construction of such models is to restrict the effective cutoff so only one replica of any given atom lies within the effective interaction radius of any other. Then, the minimum image convention can be used while computing derivatives only with respect to positions within the simulation cell. However, as we will see, this solution is inefficient, leading to quadratic scaling of the computational cost.

We now transition to the notation introduced in \cref{sub:si-mlp}. Then, $U_I$ depends not only on $\nbh{I}$, but the set of neighborhoods within $\interactions$ hops on the graph.
% , which we denote as $\nbht{I}$. 
In general, it can depend on all neighborhoods, and we can therefore write the following, which highlights the dependence of the heat flux on interactions beyond the cutoff,
\begin{align}
    \Jmpnn &= \sum_{\substack{I \in \vertices \\ JK \in \edges}} \Rm_{JI}\nonumber \\
    &\qquad \times \left( \Bigl(\dur{I}{K}{J} - \dur{I}{J}{K} \Bigr) \cdot \V_J \right) \, . \label{eq:si-J_graph}
\end{align}
This form requires the explicit computation of all elements of the Jacobian $\indur{I}{J}{K}$, or equivalently $\indur{I}{J}{}$, which in turn requires $N$-fold repeated evaluations with \ad{}, leading to the quadratic scaling observed.

\subsection{Unfolded Heat Flux}
\label{sub:si-unfolded}

Conceptually, we now modify the construction of the \mlp{}: Instead of constructing the input graph $\graph$ based on positions in $\Rsc$ only, we instead start from an extended simulation cell $\Runf$ that includes all positions up at most $\effcutoff$ distant from the atoms in $\Rsc$, i.e., all positions that contribute to $U$. The atom-pair vectors that appear in the resulting graph are identical to the standard construction, and $U$ is therefore unchanged. Similar ideas have been employed previously to define stress and heat flux under periodic boundary conditions~\cite{tpm2009t,khc2012t} without the consideration of \ad.

The unfolded heat flux $\Junfolded$ can then be obtained starting from \cref{eq:si-hf_general}.
We first write out the atom-pair vector $\R_{JI}$,
\begin{align}
    \Jpot 
    &= \sum_{\substack{I \in \Rsc \\ J \in \Runf}} 
    \left(\left[\R_I - \R_J \right] \left(\dur{I}{J}{} \cdot \V_J\right) \right) \\
    &= \sum_{\substack{I \in \Rsc \\ J \in \Runf}} 
    \left(\R_I \left(\dur{I}{J}{} \cdot \V_J\right) \right) \\
    &\quad- \sum_{\substack{I \in \Rsc \\ J \in \Runf}} 
    \left(\R_J \left(\dur{I}{J}{} \cdot \V_J\right) \right) \, .
\end{align}
In order to efficiently implement this expression with \ad{}, we must rewrite it in the form of derivatives of scalar, or low-dimensional, quantities. 

For the first term, this requires the ability to move the partial derivative outside the sum over $I$, which we enable by formally replacing the $\R_I$ appearing as pre-factors with auxiliary positions $\R^{\text{aux}}_I$ that are numerically identical, but not included in derivatives.

We can then introduce the potential energy barycenter
\begin{equation}
  \Bary = \sum_{I \in \Rsc} \R^{\text{aux}}_I U_I \, ,
\end{equation}
the first term becomes
\begin{align}
  \sum_{\substack{I \in \Rsc \\ J \in \Runf}} 
    \left(\R^{\text{aux}}_I \left(\dur{I}{J}{} \cdot \V_J\right) \right) &=
  \sum_{J \in \Runf} \frac{\partial \Bary}{\partial \R_J} \cdot \V_J  \, . 
\end{align}
The vector-vector product is taken between the denominator and the $\V_J$, not with $\Bary$.

In the second term, the sum over $I$ can be executed directly, leaving only derivatives of the scalar energy $U$ with respect to all positions in $\Runf$. So:
\begin{align}
    \Jint
    &=  \sum_{J \in \Runf} \frac{\partial \Bary}{\partial \R_J} \cdot \V_J  \\
    &\quad- \sum_{J \in \Runf} \R_J \left(  \frac{\partial U}{\partial \R_J} \cdot \V_J \right)
    \, .
\end{align}

The first term requires one backwards pass through the computation of $\Bary$ for each of its three cartesian components, or, if forward-mode \ad{} is available, a single forward pass.
The second term can be simply computed from the gradient of $U$, which is obtained in a single backwards pass.
Therefore, the asymptotic runtime scales with the number of atoms $N_{\text{unf}}$ in $\Runf$.
The number of additional atoms to be considered scales with the surface of the simulation cell, constituting a shell of width $\effcutoff$. Assuming constant density, $V$ and $N$ are proportional. Considering a cubic system for simplicity, the surface area of the simulation cell is proportional to $L^2$, where $L=V^{1/3}$. Therefore, the surface area, and the additional number of atoms, is proportional to $N^{2/3}$.

Asymptotically, this is dominated by the number of atoms $N$ in the simulation cell itself, rendering this approach formally linearly scaling in $N$.
However, for fixed $N$, the number of additional positions increases cubically with $\interactions$, restricting this approach to moderate $\interactions$.

\section{Heat Flux Benchmark}
\label{sub:si-timing}

The timings of different implementations of the heat flux were taken for ten steps of molecular dynamics of different supercells of tetragonal zirconia, using the $\interactions=2$ SchNet with $\cutoff = \SI{5}{\angstrom}$ employed for all presented results. SchNet is implemented in SchNetPack~\cite{sktm2018q}, but used with a custom calculator. The reported runtimes exclude the calculation of neighborlists, which can be cached, but include the construction of the unfolded cell. Implementations and timing scripts are available at \href{https://doi.org/10.5281/zenodo.7767432}{doi:10.5281/zenodo.7767432}.

\section{Training and Model Selection}

\subsection{Data}

Training and validation data was generated using molecular dynamics in the $NpT$ ensemble, heating a tetragonal 96-atom supercell from \si{0}{K} to a grid of target temperatures up to \si{3000}{K} at intervals of \si{750}{K} at ambient pressure, spanning the entire temperature range up to the melting point. In total, \num{10000} single-point calculations were performed using FHI-aims~\cite{FHI-aims}, with the PBEsol~\cite{przb2008t} functional, $2\times2\times2$ $k$-points, and light basis sets, with an additional basis function for \ch{O}, following the approach of Ref.~\cite{crs2017t}. The calculations were set up and run using FHI-vibes~\cite{FHI-vibes}. We use the first \num{2000} steps of each trajectory for training, and the remaining \num{500} steps for validation.

As test data, we obtained the calculations by Carbogno~\etal{}~\cite{crs2017t-data} from the NOMAD repository~\cite{NOMAD}, consisting of three trajectories of about \num{15000} steps per temperature (\si{300}{K} to \si{2400}{K} in intervals of \si{300}{K}), using every tenth step of all trajectories. This data was not used during training.

\subsection{Model and Training}

We use the SchNet~\cite{sktm2017q,sstm2018q} \mpnn{} architecture implemented in SchNetPack~\cite{sktm2018q} with varying cutoff radius $\cutoff$ and interaction depth $\interactions$, \num{128} atom features, \num{128} filter width, and a two-layer ($128 \rightarrow 64 \rightarrow 1$) output network.

For training, we use the potential energy, the forces on each atom, and the total stress tensor in a joint squared error loss function, with weights \num{0.001}, \num{0.999} and \num{100.0} respectively, and the ADAM optimizer~\cite{kb2014m}. Whenever a plateau of the loss on the validation set is encountered (with a patience of $20$ epochs), the learning rate is reduced by a factor of $1/4$ from a start of \num{e-4} to a minimum of \num{e-6}. If the loss has not improved for \num{200} epochs, the training is terminated.
Training was performed on two Tesla Volta V100 32GB GPUs using a batch size of \num{100}. 

Hyperparameters other than cutoff radius and number of interactions (see next section) were selected based on successful termination of training and minimal error on the validation set, with a final selection based on the error on the test set.

\subsection{Model Convergence with Cutoff and Interaction Depth}

We discuss the selection of the main model parameters, the cutoff radius $\cutoff$, which determines the amount of local information available, and the number of message passing steps $\interactions$, which controls the range of interactions within the network, leaving choice of training data and method, as well as model architecture fixed. 

\begin{figure}
  \centering
  \includegraphics[scale=0.6]{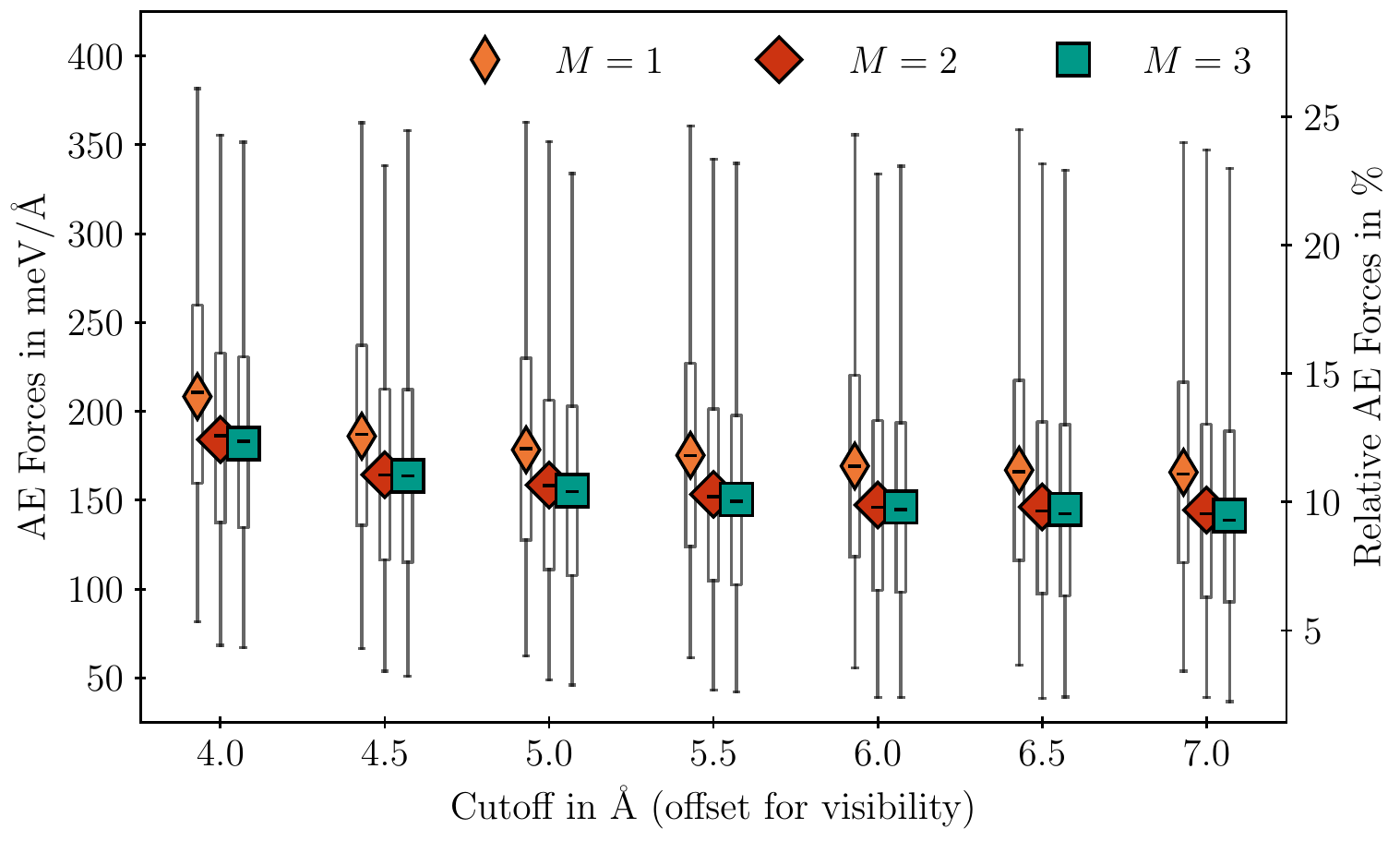}
  \caption{Absolute prediction errors (AE) for the forces on the test set, for models with different cutoffs $\cutoff$ and numbers of message passing steps $\interactions$. A constant horizontal offset has been added to $\cutoff$.
  Boxes, whiskers, bars, markers show interquartile range, total range, median, and mean, respectively.
  The relative AE is scaled by the standard deviation of forces over the test set.
  % $\cutoff = \SI{2.5}{\angstrom}$ is the radius of the first-neighbor shell for all element combinations. 
  }
  \label{fig:test_set_errors}
\end{figure}

\subsubsection{Losses}

Test set errors were evaluated for the NOMAD reference data (\cref{fig:test_set_errors}). Going from $\interactions{=\,}1$, i.e., no message passing, to $\interactions{=\,}2$ yields an approximately constant decrease in error. Additional message passing steps yield only marginal improvements. Predictive accuracy increases with cutoff radius, but saturates as the diameter of local environments exceeds the diameter of approximately \SI{10}{\angstrom} of the simulation cells used in training, approaching an all-to-all model. In this regime, all degrees of freedom are seen in the simulation cell, and message-passing cannot propagate additional information. However, $\interactions{=\,}1$ is equivalent to a non-linear pair potential, while $\interactions{>\,}1$ can model higher-order interactions, leading to lower errors. We therefore proceed with $\interactions{=\,}2$ in the following, which is sufficient to demonstrate the effect of message passing for heat flux predictions and minimises additional computational cost.

\begin{figure}
  \centering
  \includegraphics[scale=0.6]{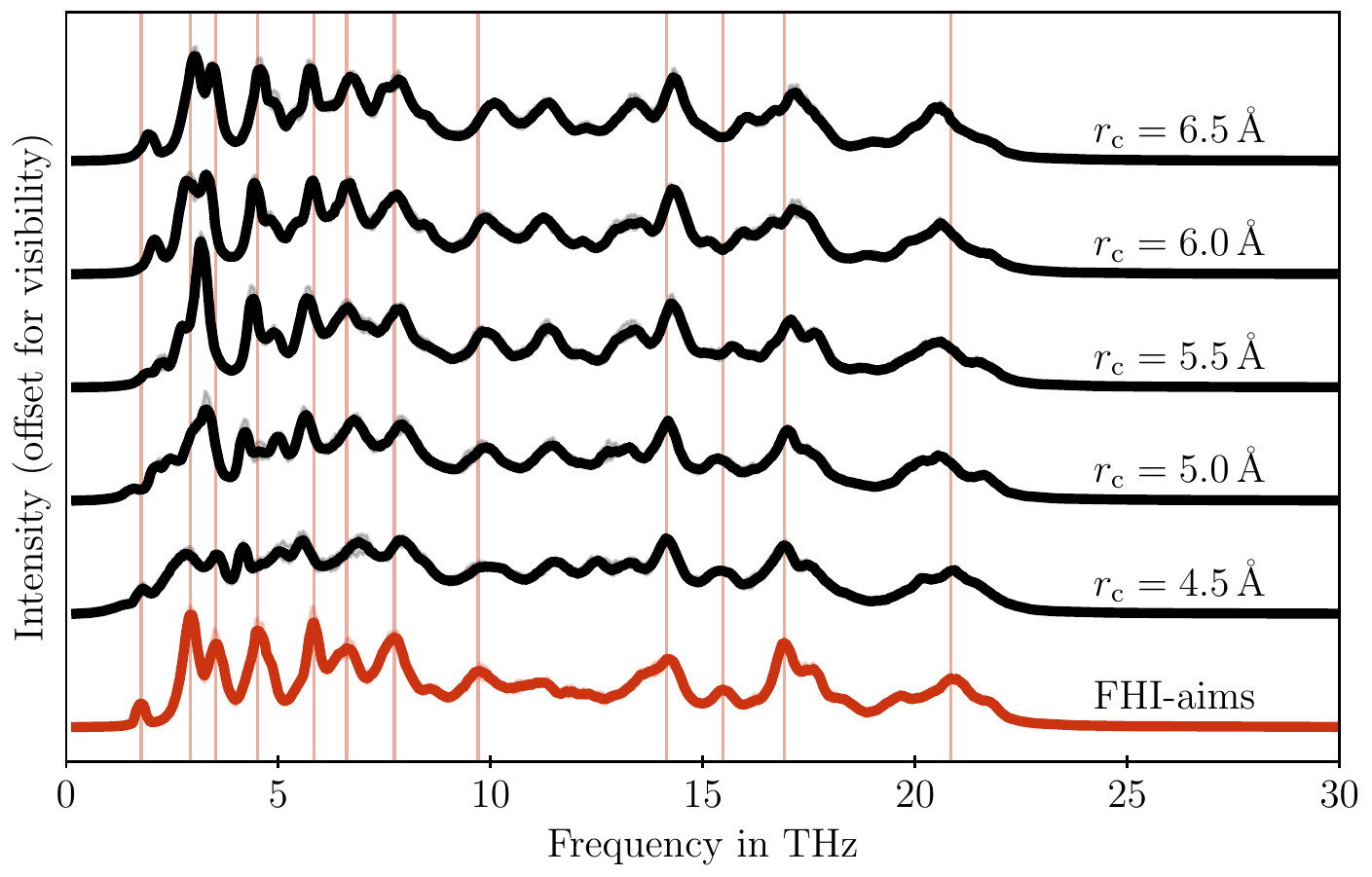}
  \caption{Comparison of vibrational density of states for \mpnns{} ($\interactions=2$) with different cutoff radii compared to a baseline computed with FHI-aims. Vertical lines indicate peaks in the FHI-aims result. Constant vertical offsets are applied to distinguish curves. Results are averaged over three trajectories of \si{60}{ps} ($\Delta t = \SI{4}{fs}$), at \SI{300}{K}, with matching initial configurations. Shaded areas indicate the minimum and maximum.}
  \label{fig:vdos}
\end{figure}

\subsubsection{Vibrational Density of States}

Since predictive accuracy on a fixed test set cannot fully predict model performance for practical applications, where larger regions of the potential energy surface are explored, we evaluate \emph{dynamical} properties as well. \Cref{fig:vdos} shows the vibrational density of states (VDOS) for different choices of $\cutoff$, evaluated for trajectories of tetragonal zirconia at \si{300}{K}, started from identical initial configurations. For $\cutoff = \SI{5}{\angstrom}$ and higher, the reference VDOS from fully ab initio \md{} is adequately reproduced. Further increasing the cutoff only leads to marginal improvements.

We therefore choose $\cutoff{=}\SI{5}{\angstrom}$ and $\interactions{=\,}2$ to compute thermal conductivities. This model is used for all results presented in this work, and used across phases.

\subsection{Testing and Limits}

% \marcomment{This section is an early draft, and is missing references, and also needs to be polished significantly.}

\begin{figure}
  \centering
  \includegraphics[scale=0.6]{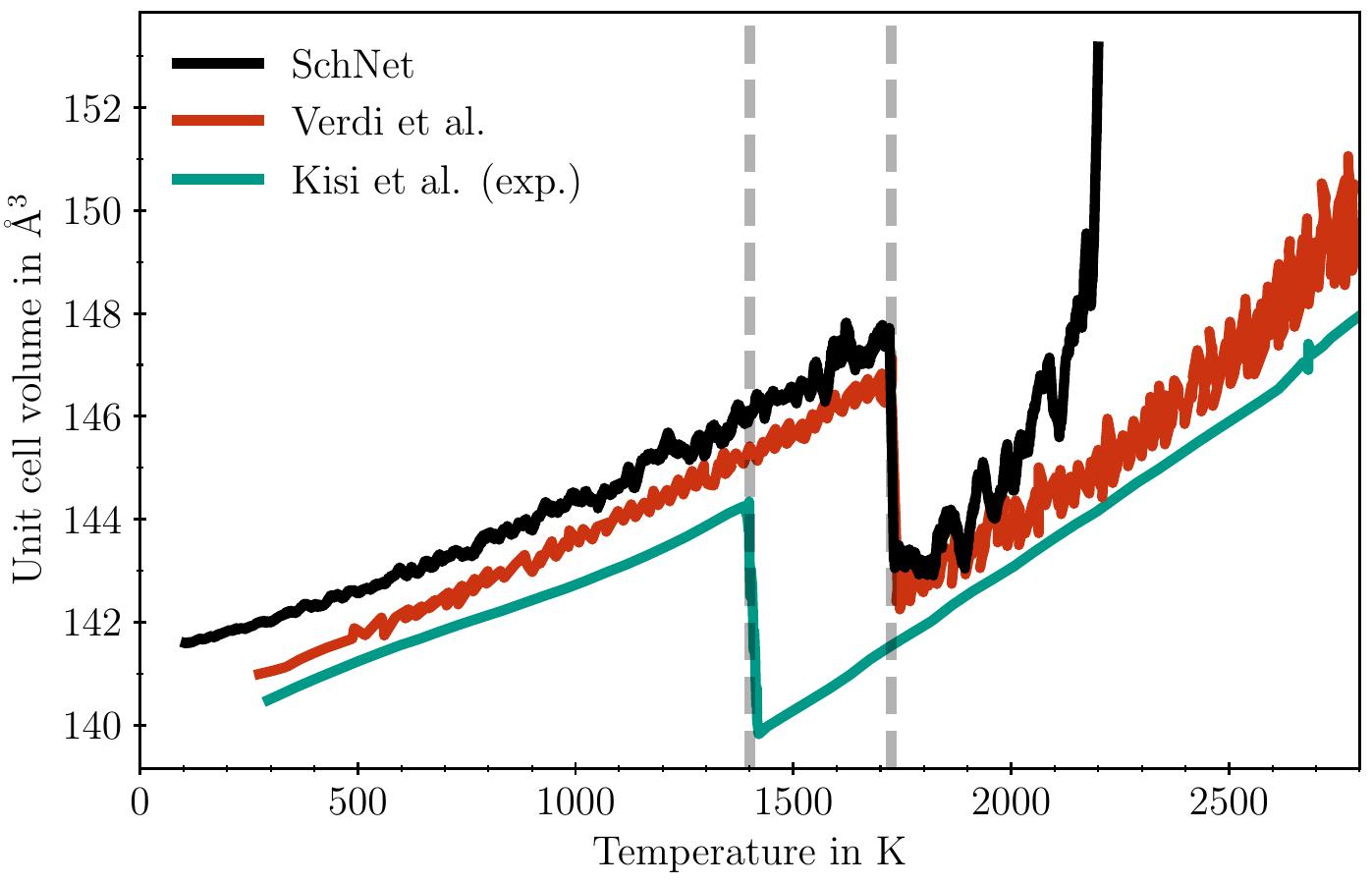}
  \caption{
  Unit cell volume, shown as rolling average over \SI{10}{ps}, versus externally imposed temperature, compared to a similar \mlp{} simulation~\cite{vkjk2021q} and experimental reference values~\cite{kh1998t}.
  Vertical line indicates the estimated transition temperatures.}
  \label{fig:si-vol_vs_temp}
\end{figure}

In this work, we employ a simple training strategy purely based on \emph{ab initio} \md{} $NpT$ simulations, as opposed to more involved schemes that rely on active learning. This approach decouples the training data from the model, allowing the straightforward comparison of different model parameters, as shown in \cref{fig:test_set_errors,fig:vdos}, but cannot be expected to yield a \pes{} for the entire phase diagram of zirconia.
In this section, we therefore probe the limits of our \mlp{} to assess whether it is sufficient for the purposes of the present work.

As a stress test, we investigate the temperature dependence of the unit cell volume, probing the monoclinic to tetragonal phase transition of zirconia, which occurs at around \SI{1480}{K} and is accompanied by a discontinuous change in volume~\cite{kh1998t}.
To this end, we perform a $NpT$ simulation with Martyna-Tobias-Klein barostat~\cite{mtk1994p} with $\tau = \SI{5}{ps}$ and stochastic velocity rescaling thermostats~\cite{bdp2007p} with $\tau = \SI{5}{ps}$ for the positions and $\tau = \SI{3}{ps}$ for the lattice, with a heating rate of \SI{1}{K \per ps}, a simulation timestep of $\SI{1}{fs}$, and a simulation cell of $324$ atoms. We compare with a similar simulation by Verdi~\etal{}~\cite{vkjk2021q}, which uses a heating rate of $\SI{0.5}{K\per ps}$.
We note that this simple approach cannot be expected to produce a quantitative estimate of the transition temperature, which require thermodynamic integration~\cite{vkjk2021q}. It can, however, indicate whether the transition is captured at all.

\Cref{fig:si-vol_vs_temp} shows the result: Despite not being explicitly trained for this task, the model qualitatively captures the phase transition, at a similar temperature to another \mlp{}. However, it over-estimates unit cell volume by approximately $\SI{1}{\percent}$, and becomes unstable above \SI{2000}{K}, likely due to insufficient training data around the tetragonal to cubic transition.
During the \gk workflow used in this work, we found that one out of eleven trajectories of \SI{1}{ns} each becomes unstable at \SI{1900}{K}; at \SI{2000}{K}, no stable \md is possible with the given potential.
Below \SI{1900}{K}, we observe no instabilities in the potential in tetragonal and monoclinic cells, running hundreds of nanoseconds of \md{} at varying supercell sizes.
% , despite running hundreds of nanoseconds of \md{} simulations.

\begin{figure}
  \centering
  \includegraphics[scale=0.6]{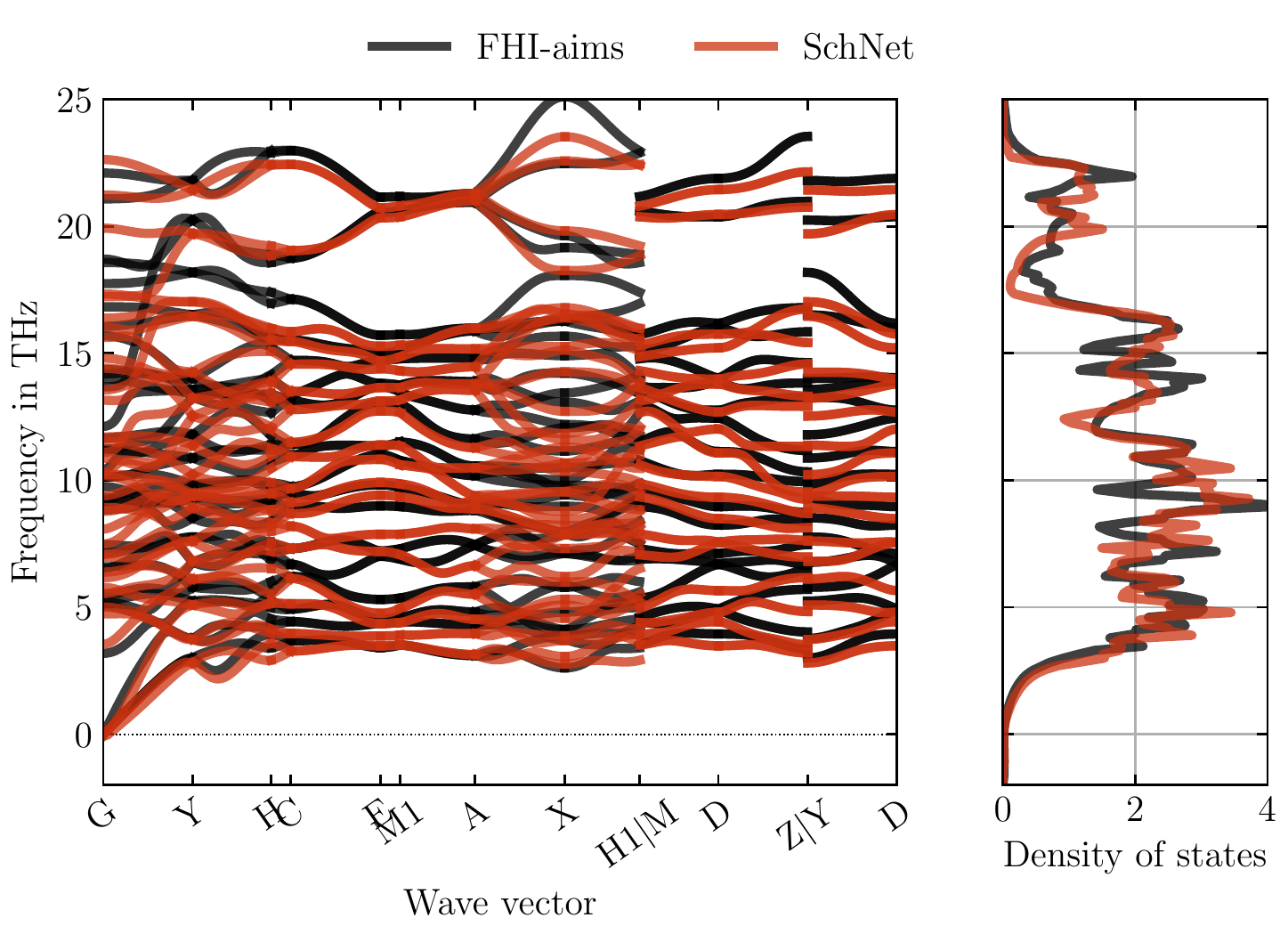}
  \caption{
  Phonon band structure and density of states for monoclininc phase of zirconia, comparing SchNet and FHI-aims.
  }
  \label{fig:si-phonons_mono}
\end{figure}

\begin{figure}
  \centering
  \includegraphics[scale=0.6]{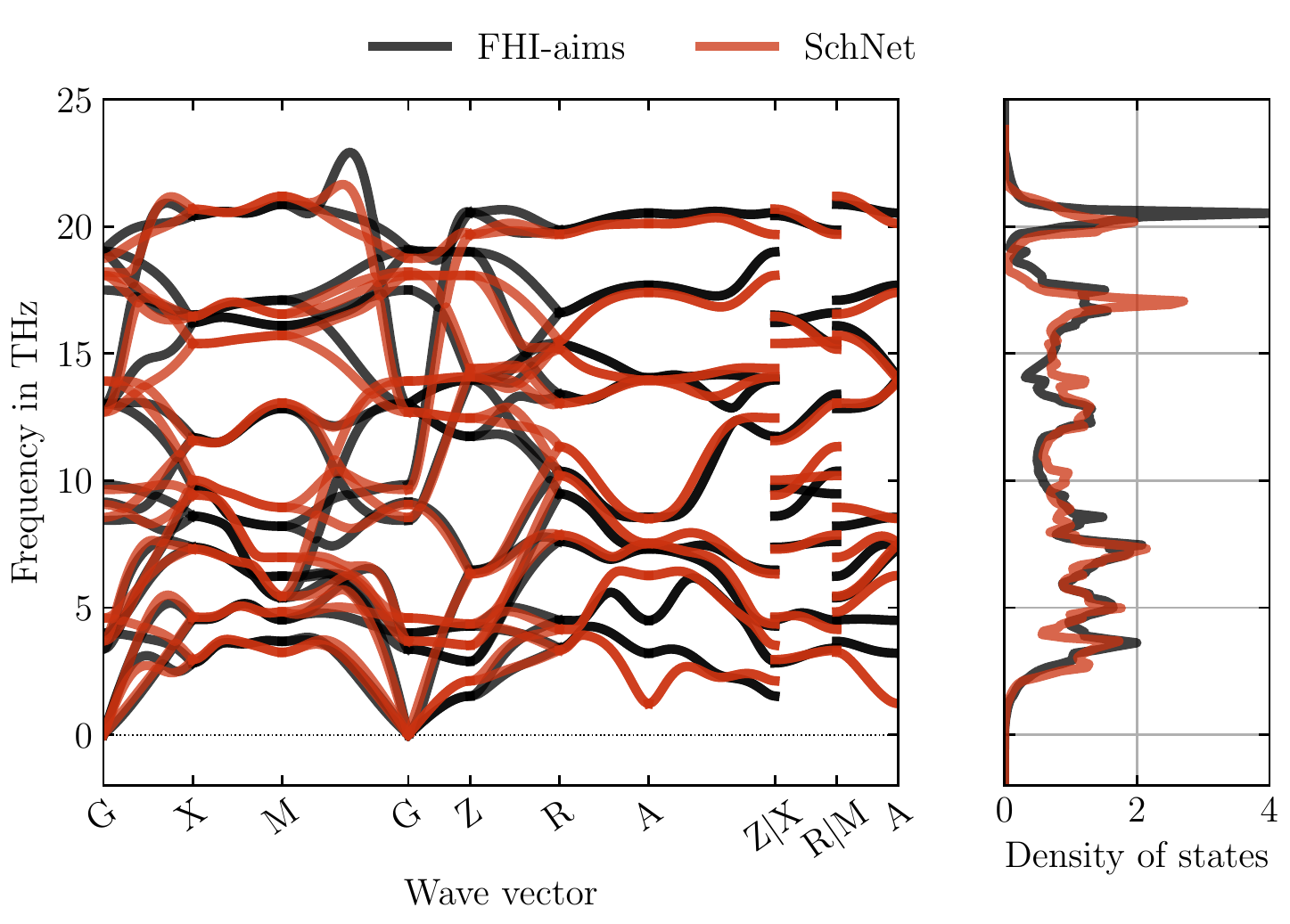}
  \caption{
  Phonon band structure and density of states for tetragonal phase of zirconia, comparing SchNet and FHI-aims.
  }
  \label{fig:si-phonons_tetra}
\end{figure}

Keeping these limitations in mind, we proceed to an investigation of the phonon band structure for the monoclinic and tetragonal phases. The results are displayed in \cref{fig:si-phonons_mono,fig:si-phonons_tetra}, with the \mlp{} adequately reproducing the \dft{} results. Supercell size convergence was checked, and we use a cell with $N=324$ atoms for these figures.
We can conclude that, while the model is limited in its ability to describe high-temperature behavior and lattice constants, it is adequate to model dynamics at lower temperatures, and therefore thermal conductivity for the temperature range investigated in the present work.

\section{Green-Kubo Convergence}

This section is concerned with the determination of overall settings and parameters for the \gk{} results presented in this work. Overall, we follow the approach presented by Knoop~\etal{}~\cite{kcs2022t}.

\subsection{General Workflow}

We use experimentally determined lattice parameters as presented by Verdi~\etal{}~\cite{vkjk2021q}, based on references~\cite{ps1969t,kh1998t}, in a 12-atom primitive cell.
From these primitive cells, simple $n\times n\times n$ supercells are created.
We proceed similarly for the tetragonal phase, with a 12-atom conventional cell in place of the 6-atom primitive cell.

The systems are then thermalised in the $NVT$ ensemble for \SI{200}{ps} at $\Delta t = \SI{4}{fs}$, and 11 decorrelated frames are extracted from the resulting trajectories, taking every \num{1000}-th step from the end.

These frames serve as the initial time-steps of the ensemble of trajectories used for the \gk{} method. We use a \SI{4}{fs} timestep throughout, consistent with previous work. The duration of the trajectories used, as well as the time-step with which the heat flux is computed, is discussed in the following subsection.

For consistency with the reference data, we report the spatial average of the thermal conductivity, $\kappa = \sum_\alpha \kappa^{\alpha \alpha}/3$. We use the noise reduction scheme introduced in Ref.~\cite{kcs2022t}, discussed further in \cref{sub:si-gk}, throughout. We use the heat flux for solids $\Jint$, which is, however, equivalent to the full $\J$, as shown in \cref{fig:si-heat_flux_convective}.

\subsection{Size and Time Convergence}

Formally, the Green-Kubo relation applies in a number of limits: infinite integration times, the bulk limit for system size, and the ensemble average over all realizations of the system at a given temperature and pressure~\cite{kcs2022t}.
% In practice, only finite simulation times and supercell sizes can be achieved, and the ensemble average is approximated by averaging over multiple trajectories in the $NVE$ ensemble with uncorrelated initial conditions sampled from an $NVT$ thermalization~\cite{kcs2022t}. 
We investigate convergence with respect to simulation time $t$ and supercell size $N$; finite ensemble size is taken into account to compute error bars.

\begin{figure}
  \centering
  \includegraphics[scale=0.6]{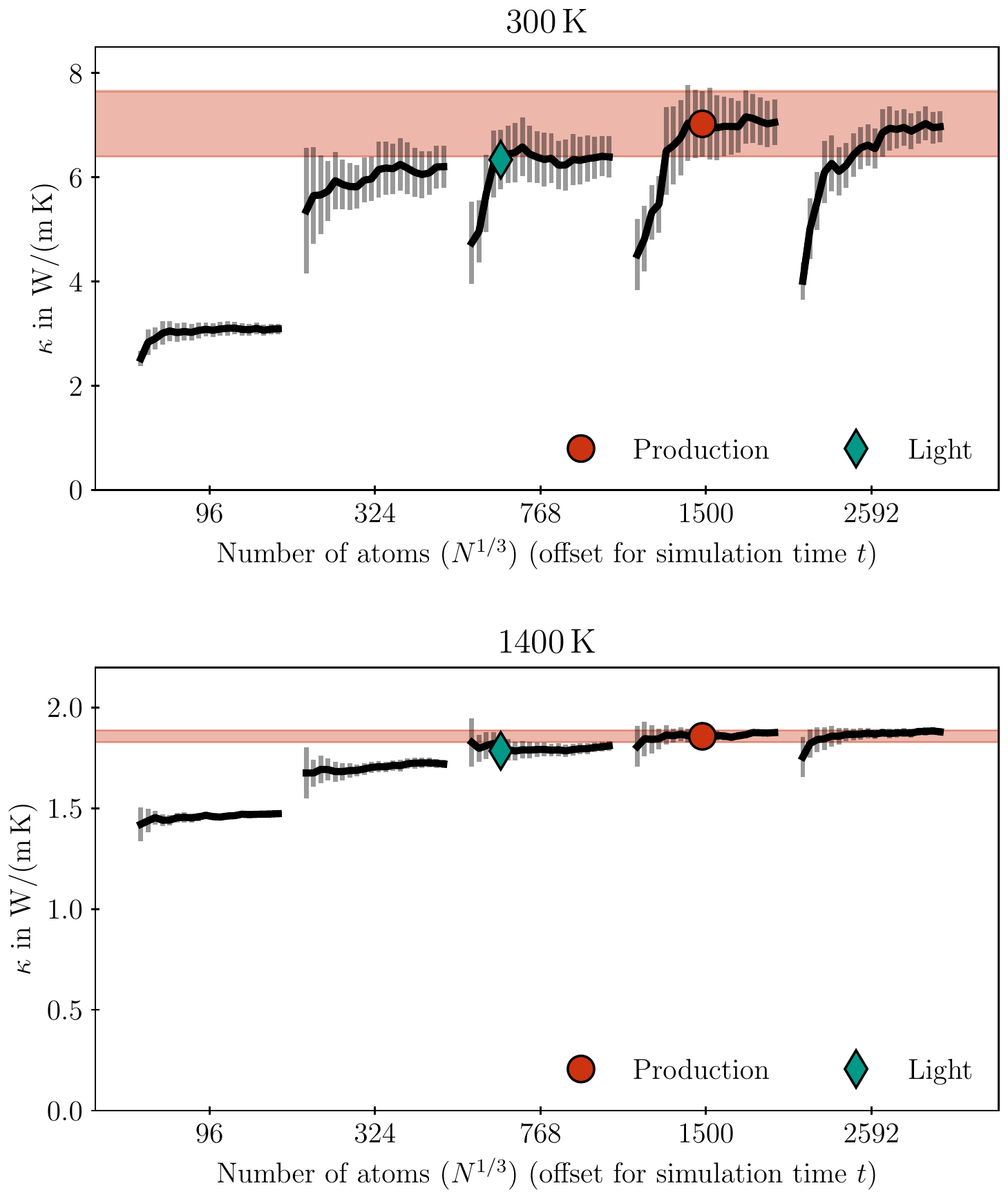}
  \caption{Convergence of thermal conductivity with respect to simulation cell size $N$ and simulation duration $t$ at \SI{300}{K} in the monoclinic phase (top) and \SI{1400}{K} in the tetragonal phase (bottom).
  For each $N$, different $t$ from \SI{0.1}{ns} to \SI{2}{ns} in steps of \SI{0.1}{ns} are shown with a constant horizontal offset. Error bars indicate the standard error.
  The x-axis is scaled to indicate the effective length scale of the simulation cell, which is proportional to $N^{1/3}$. 
  The chosen \scare{production} settings, $N{=}1500$, $t{=}\SI{1}{ns}$, as well as the \scare{light} settings, $N{=}768$, $t{=}\SI{0.5}{ns}$ are indicated.
  % , which are used for comparison of quadratically-scaling heat flux formulations, are indicated.
  For the \scare{production} settings, the standard error is shown as a shaded area.
  }
  \label{fig:kappa_convergence}
\end{figure}

% \paragraph*{Outcome} 
\Cref{fig:kappa_convergence} shows convergence with respect to simulation duration $t$ and simulation cell size $N$ at \SI{300}{K} and \SI{1400}{K}. At \SI{300}{K}, for a given $N$, simulation times of at least \SI{1}{ns} are required to obtain results that consistently lie within standard error of each other. Choosing $t=\SI{1}{ns}$, we observe no systematic increase in $\kappa$ above $N=1500$, as all results lie within the standard error of that estimate, highlighted in \cref{fig:kappa_convergence}.
At \SI{1400}{K}, anharmonic contributions dominate, leading to shorter phonon lifetimes, achieving convergence earlier.
We therefore conclude that the parameters $t=\SI{1}{ns}$ and $N=1500$ are sufficient to obtain a converged estimate for $\kappa$ at all temperatures of interest. 

For the comparison of heat flux formulations with high computational cost, we additionally require a reduced, \scare{light}, choice of parameters that yields a reasonable estimate of the converged $\kappa$. We choose $t=\SI{0.5}{ns}$ and $N=768$, which are used for the comparison of heat flux variations in Fig.~2 of the manuscript.

\subsection{Filter Frequency and Heat Flux Timestep}
\label{sub:si-gk}

% \vspace{\baselineskip}
\begin{figure}
  \centering
  \includegraphics[scale=0.6]{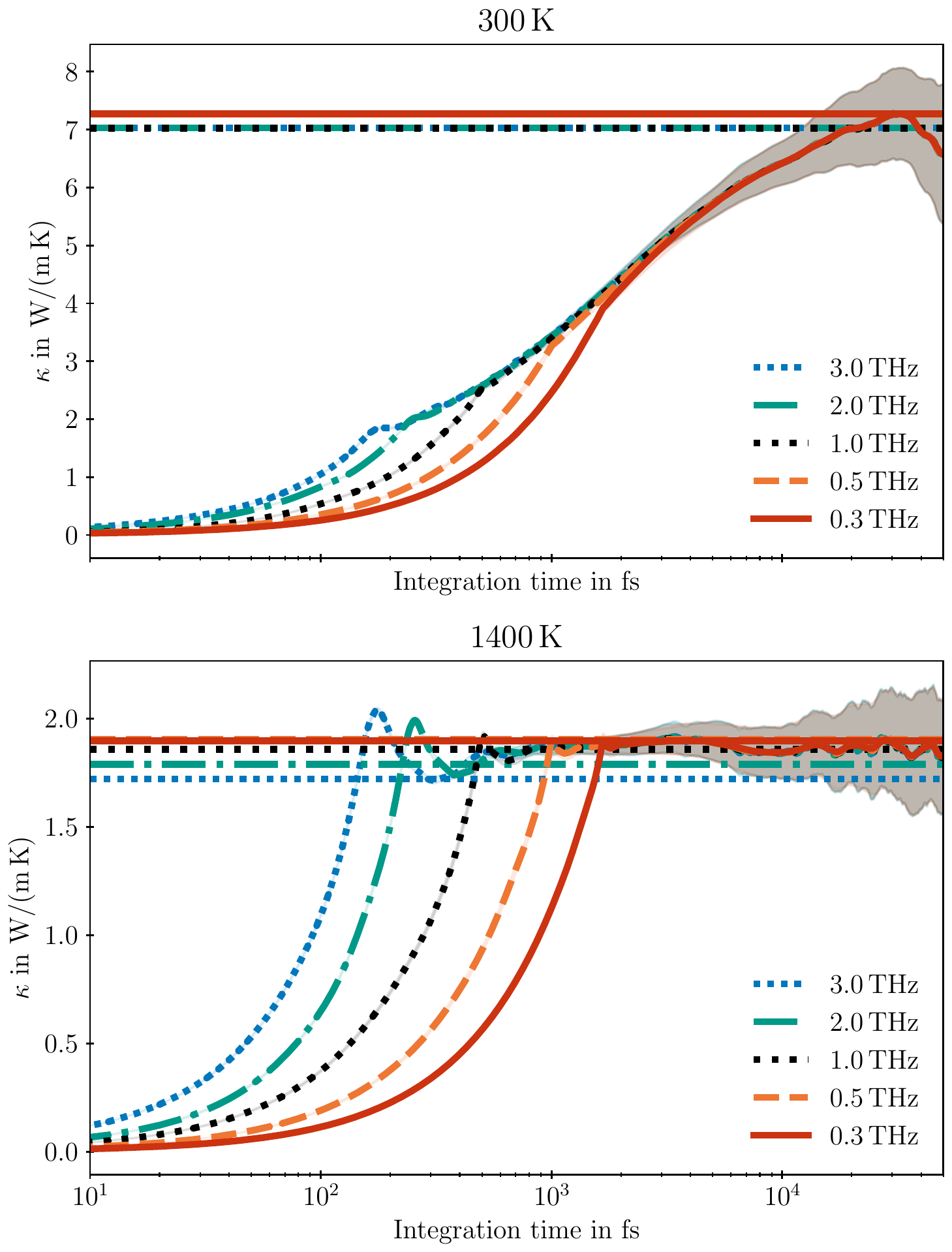}
  \caption{Comparison of the integral of the heat flux autocorrelation function for different choices of $\omega_\text{window}$, at \SI{300}{K} in the monoclinic phase (top) and \SI{1400}{K} in the tetragonal phase (bottom).
  Horizontal lines indicate chosen value for $\kappa$.
  \scare{Production} settings ($N{=}1500$, $t{=}\SI{1}{ns}$) are used.
  Shaded areas indicate standard error.
  Lower values of $\omega_\text{window}$ correspond to more severe filtering.
  }
  \label{fig:si-gk_freq}
\end{figure}

\begin{figure}
  \centering
  \includegraphics[scale=0.6]{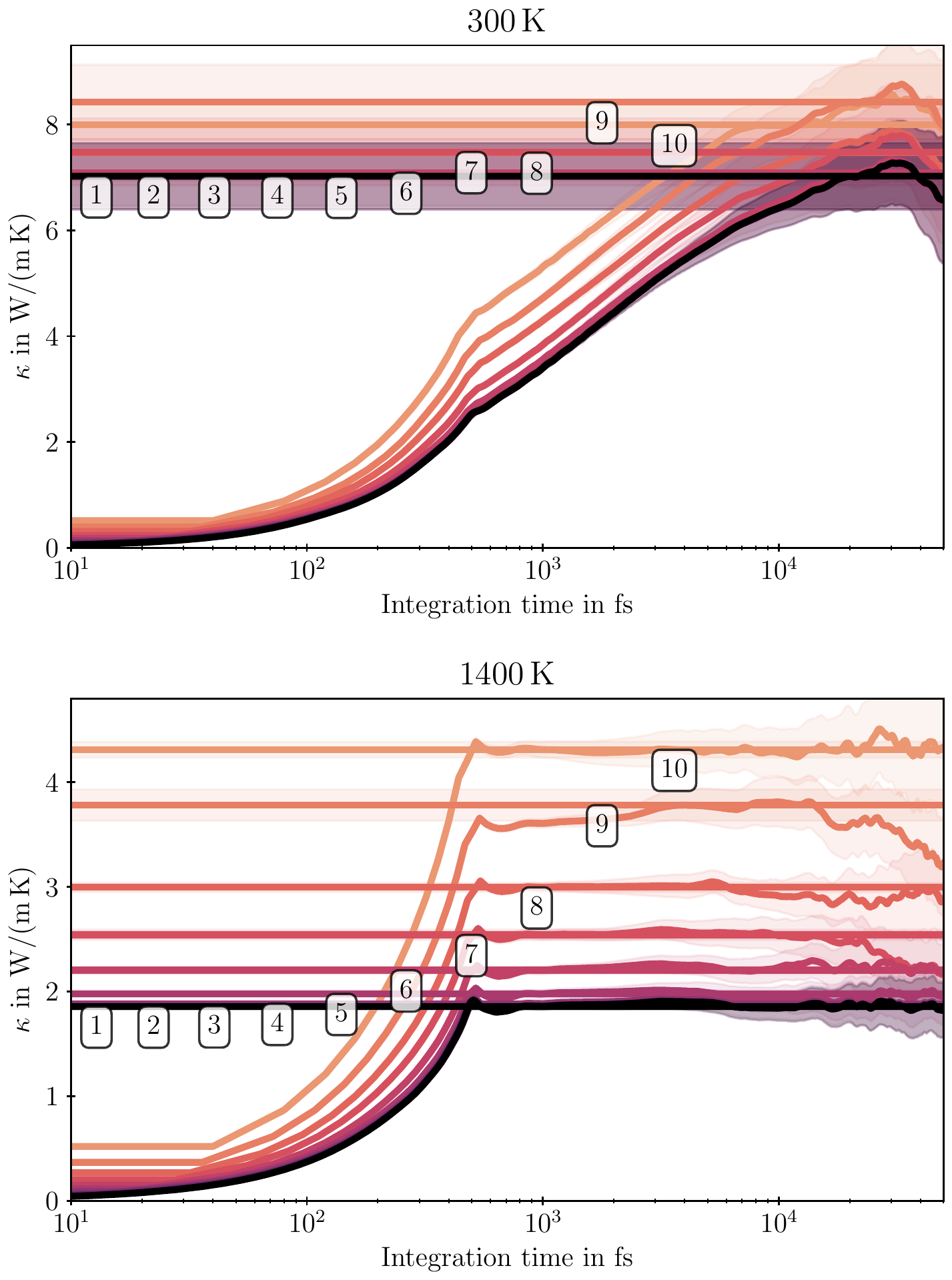}
  \caption{Comparison of the integral of the heat flux autocorrelation function for different choices of $n_{\text{hf}}$, at \SI{300}{K} in the monoclinic phase (top) and \SI{1400}{K} in the tetragonal phase (bottom).
  The upper edges of the labels indicating $n_{\text{hf}}$ are aligned with horizontal lines indicating the value for $\kappa$ chosen by the first dip of the corresponding heat flux autocorrelation function.
  \scare{Production} settings ($N{=}1500$, $t{=}\SI{1}{ns}$) are used.
  Shaded areas indicate standard error.
  }
  \label{fig:si-gk_spacing}
\end{figure}

In addition to the number of atoms in the simulation cell $N$, and the simulation duration of the trajectories used, auxiliary parameters are present in the \gk{} method as used in this work. Since the processes determining the thermal conductivity typically occur on longer timescales than the simulation timestep $\Delta t$, the heat flux can be computed at some larger timestep $n_{\text{hf}} \cdot \Delta t$ to reduce computational cost.

Additionally, the heat flux is post-processed to reduce noise and enable the automatic determination of the cutoff for the integral in the \gk{} relation. We use the noise reduction scheme presented in~\cite{kcs2022t}, consisting of two separate mechanisms: Firstly, a gauge term of the form $\sum_I \boldsymbol{\Omega}_I \cdot \V_I(t)$ is subtracted from the heat flux at every step, with the $3\times3$ matrices $\boldsymbol{\Omega}_I$ representing an average or reference atomic virial. Since our heat flux formulation does not lend itself to the computation of per-atom virials, we use virials computed for the pristine supercell. For sufficient simulation times, we find that this approximation yields satisfactory results. 
Secondly, the integrated \hfacf{} is filtered to reduce high-frequency noise, with a filter width $t_{\text{window}} = 1/\omega_\text{window}$.

In this section, we discuss the impact of the spacing parameter $n_{\text{hf}}$, and of the filter frequency $\omega_\text{window}$. We note that in principle, the impact of these settings depends on the choice of $N$ and $t$. We have, however, checked that the results presented here are robust to changes in $N$ and $t$. In general, higher $N$ and $t$ reduce noise in the \hfacf{} and therefore allow larger $\omega_\text{window}$ and higher $\Delta t$. 

\Cref{fig:si-gk_freq} shows the dependence of the integrated \hfacf, as well as the resulting\footnote{Following the approach of Ref.~\cite{kcs2022t}, we select a cutoff time based on the first zero of the \hfacf{}.} $\kappa$, on the choice of $\omega_\text{window}$.
% In this, we aim to identify the least severe choice of filter that is consistent with 
At \SI{300}{K}, no strong dependence of $\kappa$ on $\omega_\text{window}$ is observed; results lie within the standard error.
At \SI{1400}{K}, high choices of $\omega_\text{window}$ lead to misidentification of the plateau in the integrated \hfacf. From \SI{1}{THz}, further reducing $\omega_\text{window}$ yields only minor changes. We therefore choose $\omega_\text{window} = \SI{1}{THz}$.

\Cref{fig:si-gk_spacing} shows the dependence on the spacing $n_{\text{hf}}$. Across temperatures, up to $n_{\text{hf}}=3$ yields identical results. We therefore choose $n_{\text{hf}}=2$ to ensure consistent results.

\bibliography{babel_short}